\documentclass[aps,prd,onecolumn,showpacs,superscriptaddress,nofootinbib,preprintnumbers]{revtex4-2}
\usepackage{amsmath}
\usepackage{amssymb}
\usepackage{color}
\usepackage{graphicx}
\usepackage{tikz-feynman}
\usepackage{style}
\usepackage[normalem]{ulem} 
\newcommand{\half}{\dfrac{1}{2}}

\def\gtap{\ \raisebox{-.4ex}{\rlap{$\sim$}} \raisebox{.4ex}{$>$}\ }

\newcommand{\N}{\mathbf{O}}
\newcommand{\I}{\mathbf{1}}
\newcommand{\diag}{{\rm diag}}
\newcommand{\PMNS}{{\rm PMNS}}

\newcommand{\bea}{\begin{equation}\begin{array}{c}}
\newcommand{\eea}{\end{array}\end{equation}}
\newcommand{\ea}{\end{array}} 

\newcommand{\beq}{\begin{equation}}
\newcommand{\eeq}{\end{equation}}
\newcommand{\bad}{\begin{array}{ccc}}
\newcommand{\dmsol}{\mbox{$\Delta m^2_{\odot}$}}
\newcommand{\dma}{\mbox{$\Delta m^2_{\rm A}$ }}

\newcommand{\ba}{\begin{array}{c}}

\allowdisplaybreaks

\begin{document}
 
\title{Low-Scale See-Saw Mechanism and the Majorana Nature of Heavy Neutrinos}

\author{Takuya Okawa}
\affiliation{SISSA, International School for Advanced Studies, Via Bonomea 265, 34136 Trieste, Italy}
\affiliation{Istituto Nazionale di Fisica Nucleare (INFN), Sezione di Trieste, I-34127 Trieste, Italy}
\affiliation{IFPU - Institute for Fundamental Physics of the Universe, Via Beirut 2, 34014 Trieste, Italy}

\author{S. T. Petcov}
\altaffiliation{Also at: Institute of Nuclear Research and Nuclear Energy, Bulgarian Academy of Sciences, 1784 Sofia, Bulgaria}
\affiliation{INFN/SISSA, Via Bonomea 265, 34136 Trieste, Italy}
\affiliation{Kavli IPMU (WPI), UTIAS, University of Tokyo, Kashiwa, Chiba 277-8583, Japan}

\begin{abstract}
  We investigate the possibility to determine the Majorana nature of the heavy
  Majorana neutrinos present in the low-scale type I seesaw mechanism of
  neutrino mass generation in experiments at
  $e^+ \text{-} e^-$ and $\mu^+ \text{-} \mu^-$ colliders. This is done
  on the examples of two specific symmetry-protected seesaw scenarios with
  two right-handed
  (RH) singlet neutrinos and one seesaw realization with three RH neutrinos.
  We use the forward-backward symmetry in the angular distribution
  of the heavy neutral leptons produced 
  in the $e^+ \text{-} e^-$ and $\mu^+ \text{-} \mu^-$ collisions
  as an indicator of their Majorana nature.
  Our results indicate that the experiments at FCCee during its first phase
  of operation have a remarkable potential not only to discover
    the heavy neutral leptons associated with the type I seesaw mechanism
    if they have masses smaller than the $W^\pm$-boson mass, but also
    to probe and provide evidence for their Majorana nature.
\end{abstract}

\maketitle

\section{Introduction}
\label{sec:intr}

In the present article we investigate the possibility to determine the Majorana nature of the Heavy Neutral Leptons (HNLs) present in the low-scale type I seesaw mechanism of neutrino mass generation in experiments at $e^+ - e^-$ (FCCee, CEPC, ILC, CLIC) and $\mu^+ - \mu^-$ colliders. The type-I seesaw \cite{Minkowski:1977sc,Yanagida:1979as,GellMann:1980vs,Glashow:1979nm,Mohapatra:1979ia} is arguably the most natural extension of the Standard Model that can account for the generation and smallness of the non-zero light neutrino masses. It consists of adding to the Standard Model (SM) at least two right-handed (RH) neutrino fields, which are SM gauge symmetry singlets. If the additional assumption of total lepton charge conservation is not imposed, the RH neutrino fields admit a Majorana mass term in the Lagrangian which preserves the SM gauge symmetries. These fields also couple to the Higgs and the three left-handed (LH) lepton doublets via gauge-invariant Yukawa interactions. Following the spontaneous breaking of the electroweak symmetry, the LH and RH neutrino fields form a neutrino
Dirac mass term. The interplay between the Dirac and the RH neutrino Majorana mass terms leads to a set of seesaw-suppressed masses for the observed light neutrinos, and to heavy massive neutrinos, both of which are Majorana particles. In this setting the LH flavour (the RH) neutrino fields are linear combinations of three massive light (sub eV scale) and, in general, $k\ge 2$ heavy (much heavier than the eV scale) Majorana neutrino fields $\chi_i$ and $N_j$, $i=1,2,3$, $j =1,2,...,k$.

Another remarkable feature of the type-I seesaw extension of the Standard Model is that it provides a mechanism \cite{Kuzmin:1985mm,Fukugita:1986hr} for a dynamical production of the observed baryon asymmetry of the Universe (BAU) \cite{Planck:2018vyg, Cooke:2017cwo}, thus connecting the generation and the smallness of the non-zero neutrino masses with the generation of BAU. 

Observing the HNLs $N_j$ in experiments at colliders would be a
remarkable discovery. Establishing
experimentally that the so observed HNLs are Majorana particle(s)
would be critical evidence,
if not a proof, that they are associated with the low-scale type
I seesaw mechanism of neutrino
mass generation.
This would constitute a breakthrough in our understanding of the origin of
the light neutrino masses and
mixing, provide a proof that the total lepton charge is not conserved in Nature
and will have far-reaching fundamental implications in particle physics and
the physics of the Early Universe.

 The future prospects of heavy neutrino searches at next-generation
 lepton colliders were discussed, e.g., in \cite{Banerjee:2015gca}.
The authors focus, in particular, on the planned $e^+ - e^-$
 colliders considering various production and decay modes of the heavy neutrino
 (N). They conclude that the final state with $l^- +2j +~{\rm missing~energy}$,
 arising from the $e^+ + e^- \to N + \nu_l$ production mode,
$\nu_l$ being a flavour neutrino, $l=e,\mu,\tau$,
 is the most
 promising channel,
 but that it is insensitive to the Majorana nature of the heavy neutrino.
 The authors  study also a new production channel
 $e^+ + e^- \to N + l^\pm + W^\mp$, which leads to a final state with same-sign dileptons
 plus four jets, which allows one to  directly probe the lepton number
 violation in  $e^+ - e^-$ colliders. 
 
A comprehensive study of sterile neutrino (HNL) searches at $e^+ - e^-$,
$p - p$ and $e^- - p$ colliders was performed in \cite{Antusch:2016ejd}
\footnote{The article \cite{Antusch:2016ejd}
contains also an extensive reference list of earlier studies.}. 
In \cite{Antusch:2016ejd} a symmetry-protected seesaw mechanism with two
RH neutrinos at the TeV scale was considered.
The analysis was performed
in the limit of exact lepton charge conservation in which the
two heavy Majorana neutrinos, present in this scenario, form a Dirac fermion.
The problem of determination of the Majorana nature of the two heavy 
neutrino mass eigenstates was not addressed in  \cite{Antusch:2016ejd}.

At $p - p$ colliders (LHC) the Majorana nature of produced HNLs can manifest
itself in the existence of processes with same-sign dileptons,
$l^- +  l^-$ ($l^+ +  l^+$), $l=e,\mu,\tau$, in the final state,
in which the lepton charge $L_l$ is not conserved and changes by two units.
These processes can originate from a chain of reactions in which a virtual or
real $W^{-(+)}$ produced in the $p - p$ collisions
decays via $W^{-(+)} \rightarrow l^{-(+)} + N_j$, the real Majorana $N_j$ decaying in turn
into $l^{-(+)} + W^{+(-)}$, with the
$W^{+(-)}$ producing two jets $j^{(\prime)}_1 + j^{(\prime)}_2$ in the final state:
$p + p \rightarrow X + W^{-(+)} \rightarrow  X + l^{-(+)} + N_j \rightarrow  X + l^{-(+)} + l^{-(+)} + W^{+(-)} \rightarrow  X + l^{-(+)} + l^{-(+)} + j^{(\prime)}_1 + j^{(\prime)}_2$.
One of the signatures
of this process is that the invariant mass of one of the charged leptons and
the two jets should be equal to the mass of $N_j$, $M_j$.

Such processes have been discussed, e.g.,
in \cite{Atre:2009rg,Degrande:2016aje,Ruiz:2020cjx}, 
and searched for by the ATLAS and CMS experiments at the LHC. The production of a single heavy Majorana
neutrino $N$ has been considered. The searches by ATLAS and CMS gave negative results and
combined limits on the mass of $N$, $M_N$, and its couplings to the electron and muon
in the weak charged current,
$\Theta_{eN}$ and $\Theta_{\mu N}$, have been obtained
\cite{CMS:2022fut,CMS:2024oyn,arXiv:2503.16213}. For example, at
$M_N = 10~(20)$ GeV, the CMS collaboration reported
\cite{CMS:2023hwf,CMS:2024oyn} the following upper limits:
$|\Theta_{eN}|^2 < 0.7~(4.8)\times 10^{-6} $ and
$|\Theta_{\mu N}|^2 < 0.4~(2.9)\times 10^{-6}$ at 95\% C.L.;
at larger or smaller  values of $M_N$ the limits are noticeably less
stringent. Similar limits for $M_N = 10$ GeV were reported by the ATLAS
collaboration \cite{arXiv:2503.16213}, $|\Theta_{eN}|^2 < 0.3\times 10^{-6} $
and $|\Theta_{\mu N}|^2 < 0.2\times 10^{-6}$; at $M_N = 20$ GeV the ATLAS corresponding upper limits are \cite{ATLAS:2019kpx} $\sim 10^{-5}$.

It was shown in 1984 in \cite{Petcov:1984nf} that under general conditions,
which are fulfilled, e.g., in the low-scale seesaw scenarios, 
a Majorana fermion $N$  produced via 
$e^+ + e^- \rightarrow N + f~(\bar{f})$
and  $\mu^+ + \mu^- \rightarrow N + f~(\bar{f})$, where $f$ is
an unobservable in the process spin-1/2 particle that could be, e.g., a
$\nu_l$ ($\overline{\nu_l}$), has in the $e^+ - e^-$ c.m. an angular 
distribution, which is forward-backward (F-B) symmetric
with respect to the direction of the 3-momentum of, e.g., $e^-$, ${\bf p}_{e^-}$,
while for a Dirac fermion this symmetry does not hold.
This Majorana particle ``indicator'' was rediscovered in the context of
the type I seesaw scenarios 21 years later in \cite{delAguila:2005pin} and
again 37 years later in \cite{Blondel:2021mss} and successfully
applied to a single $N$ production in $e^+ - e^-$ annihilation.
The authors of \cite{delAguila:2005pin}
considered the process of
$e^+ + e^-  \to N\nu \to lW\nu$, $l=e,\mu,\tau$, at the discussed future 
colliders CLIC with energy of 3 TeV and ILC with energy of
500 GeV and have shown that using
the F-B symmetry indicator one can probe the nature - Dirac or Majorana -
of N having a mass in the range $M_N \sim (1 - 2)$ TeV
($M_N\sim (100-400)$ GeV) at CLIC (ILC),
provided the coupling of $N$ to the electron
$|\Theta_{eN}|\gtap 0.004$ ($|\Theta_{eN}|\gtap 0.005$).
In \cite{Blondel:2021mss} the authors investigated the production of
$N$ at FCCee working at the $Z$-pole.
They showed that using the F-B symmetry
feature or absence of it
it could be possible, in principle, to observe $N$ with a mass
$M_N \sim (10 - 70)$ GeV and determine whether it is a Dirac fermion or
a Majorana particle. Similar conclusion based on the results obtained
in \cite{Petcov:1984nf} was reached 
in \cite{arXiv:2403.06561} where the authors discussed the production
of a single seesaw-like neutral lepton $N$
together with $\nu_\mu$ or $\overline{\nu_\mu}$ at a $\mu^+ - \mu^-$
collider.

In the studies \cite{delAguila:2005pin,arXiv:2403.06561,Blondel:2021mss}
the production of a single HNL was considered. However, as is well known,
the type I seesaw mechanism with
one RH neutrino predicts the existence of two massless light neutrinos.
This is incompatible with the neutrino oscillation data which show that at
least two of the light neutrinos must have non-zero masses. Thus, realistic
type I seesaw scenarios must include at least two RH neutrinos, and
correspondingly, at least two heavy Majorana neutrinos (or HNLs),
$N_j$, $j=1,2,...,k$, $k\ge 2$.

In \cite{Hernandez:2018cgc} the authors,
    working with the symmetry-protected
``extended'' version of the seesaw mechanism with two RH neutrinos
\cite{Kang:2006sn} (see also, e.g., \cite{Lopez-Pavon:2012yda}),
and correspondingly two heavy Majorana neutrinos $N_{1,2}$,
which form a pseudo-Dirac pair \cite{Wolfenstein:1981kw,Petcov:1982ya},
investigated the possibility of constraining 
the pseudo-Dirac mass splitting $|M_2 - M_1|$ at the
ILC  $e^+ - e^-$ collider with energy of 250 GeV.
In the considered seesaw scenario the pseudo-Dirac mass splitting is a lepton
charge violating quantity related to the Majorana nature of the two heavy
neutrinos $N_{1,2}$ forming the pseudo-Dirac one.
As a signature of the Majorana $N_{1,2}$ production the authors used the
forward-backward symmetry indicator
in the angular distribution of  $N_{1,2}$,
and correspondingly of the charged lepton $l^-$ (or $l^+$), $l=\mu,e$,
originating from the decays $N_{1,2} \to l^- + W^+$ (or $l^+ + W^-$),
with the virtual $W^+$ ($W^-$) producing two jets in the final state.
Assuming integrated luminosity of the experiment of  2 $\mathrm{ab}^{-1}$,
the authors concluded, after performing numerical analysis in two benchmark
cases with $M_{1,2} = 5$ GeV and 20 GeV and couplings
$\Theta_{l1,2N} \sim (10^{-6} - 10^{-5})$, $l=e,\mu,\tau$,
that experiments at the ILC can probe exceptionally small
values of the $N_1 -N_2$ mass splitting
$|M_2 -M_1|\gtap 10^{-4}$ eV and  $|M_2 -M_1|\gtap 10^{-2}$ eV, respectively.

In the present article we continue the studies of 
the  possibility to determine
the Majorana nature of the HNLs, which are an integral part of the
minimal type I seesaw  mechanism of light neutrino mass generation,
in experiments at $e^+ - e^-$ (FCCee, CEPC, ILC, CLIC) and
$\mu^+ - \mu^-$ colliders.
FCCee working at the $Z^0$-pole during its first phase
  is foreseen to produce $6\times 10^{12}$ $Z^0$-bosons in four years
  of running with integrated luminosity of 205 $ab^{-1}$. This offers a
  remarkable opportunity to test and obtain evidence, if not a proof, of 
  the Majorana nature of the seesaw heavy neutrinos if they have masses
  not exceeding that of the $W^\pm$-boson.
  
 We consider two rather well known low-scale
seesaw scenarios with two RH neutrinos,
which differ from that discussed in \cite{Hernandez:2018cgc}, 
and one low-scale scenario with three RH neutrinos.
The scenarios with two RH neutrinos are of the ``symmetry-protected'' class,
in which the respective heavy Majorana neutrinos $N_{1,2}$ form a pseudo-Dirac
pair. In both cases  $N_{1,2}$ have relatively large charged (neutral)
current  weak interaction couplings with the charged leptons and
the SM $W^\pm$-bosons (flavour neutrinos and the $Z^0$-boson).
The two scenarios differ by the phenomenologically allowed
magnitude of the $N_{1} - N_{2}$ mass splitting, $|M_2-M_1|$,
which in the considered cases is a lepton charge violating quantity
directly related to the Majorana nature of $N_{1,2}$.
The seesaw scenario with three RH neutrinos
is one example of seesaw models investigated recently in
\cite{Granelli:2026qss}, in which both the charged (neutral) couplings of the
heavy Majorana neutrinos $N_{1,2,3}$ and the differences between their masses
can be relatively large and might allow their observation
in experiments at $e^+ - e^-$ ($\mu^+ - \mu^-$) colliders.
After reviewing in some detail the considered specific seesaw
realisations, we discuss next  the production of the heavy Majorana neutrinos
$N_{j}$, $j=1,...,k$, $k=2$ or 3, present in these realisations
at $e^+ - e^-$ ($\mu^+ - \mu^-$) colliders. Each $N_{j}$ 
can be produced in the process of 
$e^+ + e^-$ ($\mu^+ + \mu^-$) annihilation together with
$\bar{\nu}_e$ ($\bar{\nu}_\mu$) or $\nu_e$ ($\nu_\mu$) via virtual 
$W^+$- or $W^-$-exchange diagrams, and with $\bar{\nu}_{\ell'}$ or
$\nu_{\ell'}$, $\ell' = e,\mu,\tau$, via $Z^0$-exchange diagrams.
We assume that $N_j$ undergoes the decay  $N_j \to l^- + W^+$,
$l=e,\mu,\tau$, with virtual or real  $W^+$ producing two identifiable
jets in the final state via $W^+ \to j_1 + j_2$.
When $N_j$ is produced with $\bar{\nu}_e$ or  $\bar{\nu}_{\ell'}$,
the decay $N_j \to l^- + W^+$ is allowed independently of whether
$N_j$ is a Majorana or Dirac fermion since the corresponding chain of
processes is Lepton Number Conserving (LNC).
However, in the case of $e^+ + e^- \to N_j + \nu_e (\nu_{\ell'})$,
the same decay $N_j \to l^- + W^+$ is allowed only if $N_j$ is a Majorana
particle and is forbidden for a Dirac $N_j$ as the corresponding chain
of processes is Lepton Number Violating (LNV).
The presence of the LNV contribution in addition to the LNC one
in the differential  cross section of the process of interest,
$e^+ + e^- \to N_j + \overline{\nu_{\ell'}}~(\nu_{\ell'}) \to l^- + W^+ + E_{mis} \to l^- + j_1 + j_2 + E_{mis}$,
$d\sigma(e^+ + e^-\to l^- + j_1 + j_2 + E_{mis})/d\cos\theta$,
where $E_{mis}$ stands for missing energy,
$(p(l^-)+p(j_1)+p(j_2))^2 \equiv M^2_{l^-j_1j_2} = M^2_j$,
and $\theta$ is the angle between the 3-momenta of $N_j$ and $e^-$
in the $e^+ - e^-$ c.m.s., $p(l^-)$, $p(j_1)$ and $p(j_2)$ being the
4-momenta respectively of $l^-$, $j_1$ and $j_2$,
ensures that the $\cos\theta$-distribution of the produced $N_j$ is
fully forward-backward (F-B) symmetric. The zero F-B asymmetry,
$A_{FB} = 0$, in the production of $N_j$ is thus a signature of the Majorana
nature of $N_j$; in the absence of the LNV contribution,
the  $\cos\theta$-distribution of interest generated by the LNC contribution
would exhibit a non-zero  F-B asymmetry, $A_{FB} \neq 0$, 
which would signal  the production of a Dirac fermion.
Using these indicators we proceed to investigate the
possibility to determine the Majorana nature of the heavy Majorana neutrinos
$N_j$ present in the considered seesaw scenarios.
The aim of our study is not to evaluate the
feasibility of the discussed experiments to discover the heavy
Majorana neutrinos associated with these seesaw scenarios
\footnote{The heavy Majorana neutrinos of the seesaw mechanism 
constitute a class of Heavy Neutral Leptons (HNLs), the search for which
is one of the goals of the program of research in neutrino and particle
physics (see, e.g., \cite{Antel:2023hkf}).} 
and probe their nature (such a study will be performed elsewhere),
but rather to illustrate the new aspects and the difficulties arising in
this highly challenging problem when one takes into account the existence of
more than one heavy Majorana neutrino, directly related to the low-scale
type I seesaw mechanism of neutrino mass generation.

The present article is organised as follows.
In Section \ref{sec:seesaw} we  briefly review the low-scale type
I seesaw mechanism of neutrino mass generation.
The specific seesaw scenarios with two and three RH neutrinos
we will consider in our investigation of the possibility to 
determine the nature of the heavy Majorana neutrinos
present in these scenarios  
are discussed  in Section \ref{sec:spseesaw}.
We investigate next in detail in Section \ref{sec:DvsM} 
the production of the heavy Majorana neutrinos
$N_{j}$, $j=1,...,k$, $k=2$ or 3, present in the specific seesaw
realisations described in Section \ref{sec:spseesaw},
in $e^+ - e^-$ ($\mu^+ - \mu^-$) collisions.
Section \ref{sec:SC}
contains the summary and conclusions of the present work. 

%
\section{The Low-Scale Type I Seesaw Framework}
\label{sec:seesaw}
%

We briefly review the low-scale type I seesaw mechanism of neutrino
mass generation. This allows us to set the notation for our subsequent
analysis. We follow the discussions in
\cite{Ibarra:2010xw,Ibarra:2011xn,Dinh:2012bp,Penedo:2017knr,Granelli:2026qss}.

We consider the standard type I seesaw scenario \cite{Minkowski:1977sc,Yanagida:1979as,GellMann:1980vs,Glashow:1979nm,Mohapatra:1979ia}, in which we extend the Standard Model (SM) by adding $k = 2$ or 3 ``heavy'' right-handed (RH) neutrino fields $\nu_{aR}$, $a=1,...,k$, $k=2$ or 3. We assume that the fields $\nu_{aR}$ are Standard Model gauge singlets, have Yukawa couplings with the left-handed (LH) lepton doublet fields and possess a ``large'' Majorana mass. We work in the charged lepton mass basis \footnote{This can be done without loss of generality.}. The neutrino mass term in the Lagrangian of the considered extension of the SM is given by:
\begin{equation}
\mathcal{L}_{\nu}\;=\; -\, \overline{\nu_{\ell L}}\,(M_{D})_{\ell a}\, \nu_{aR} - \half\, \overline{\nu^{C}_{aL}}\,(M_{R})_{ab}\,\nu_{bR}\;+\;{\rm h.c.}\,, 
\label{typeI}
\end{equation}
%
where $\nu_{\ell L}$, $\ell = e,\mu,\tau$ are the LH flavour neutrino fields, $\nu^{C}_{aL}\equiv C \overline{\nu_{aR}}^T$, $C$ being the charge conjugation matrix, $M_{R} = (M_{R})^T$ is the $k\times k$ Majorana mass matrix of the RH neutrinos, and $M_{D}$ is a $3\times k$ neutrino Dirac mass matrix, which is generated by the matrix of neutrino Yukawa couplings after the electroweak (EW) symmetry breaking. The matrices $M_{R}$ and $M_{D}$ are complex, in general. The full neutrino mass matrix in Eq. (\ref{typeI}) can be set in a block diagonal form by the following transformation: 
\begin{eqnarray}
 \mathcal{M} =  \left(\begin{array}{cc}
	\N & M_D \\
	M_D^T & M_R \\        
       \end{array}
 \right) & = &
\Omega\,
 \left(\begin{array}{cc}
	U\hat{m}U^T & \N \\
	 \N^{T} & V\hat{M_N}V^T \\        
       \end{array}
\right)\,\Omega^T
\label{seesaw}
\end{eqnarray}
%
where $\Omega$ is a $(3+k)\times (3+k)$ unitary matrix,
$\hat{m}\equiv\diag(m_1,m_2,m_3)$ is a diagonal matrix with masses of the
light Majorana neutrinos $\chi_i$, $i=1,2,3$, $\hat{M_N}\equiv\diag(\hat{M_1},\hat{M_2},\ldots,\hat{M_k})$ is a diagonal matrix containing the masses $\hat{M_{j}}$ of the heavy Majorana neutrino mass eigenstates $N_{j}$, $j = 1,2,...,k$, $U$ and $V$ are unitary matrices diagonalising respectively the Majorana mass matrices $m_\nu$ and $M_R$ of the LH flavour and the RH neutrinos.

The unitary diagonalization matrix $\Omega$ can be formally expressed as the exponential of an antihermitian matrix:
\begin{eqnarray}
 \Omega & = & \exp\left(\begin{array}{cc}
  \N & R \\
  -R^\dagger & \N
\end{array}\right)\;=\;\left(
\begin{array}{cc}
\I-\half R R^\dagger & R \\
-R^\dagger & \I-\half R^\dagger R
\end{array}\right)\;+\;\mathcal{O}(R^3)\,,
\label{R0}
\end{eqnarray}
where $R$ is a $3\times k$ complex matrix and the second equality is obtained assuming that $R$ is ``small''.

Diagonalising the neutrino mass matrix given in Eq. (\ref{seesaw}) leads, as we have noticed, to light and heavy Majorana neutrinos $\chi_{iL}$ and $\hat{N}_{jL}$ with masses $m_i$ and $\hat{M_j}$, whose
fields satisfy the Majorana condition:
\begin{equation}
  C\, (\bar{\chi_i})^T = \chi_i\,,\quad
 C\, (\overline{\hat{N}_j})^T = \hat{N}_j\,, ~~i=1,2,3,~~j=1,2,..., k\,.
 \label{eq:Majcond1}
\end{equation}
%
The LH flavour neutrino fields are linear combinations of the LH components
of the light and heavy Majorana neutrino fields $\chi_{iL}$ and $\hat{N}_{jL}$:
\begin{equation}
  \nu_{lL} =
  \left( \left ( \I - \dfrac{1}{2}R\,R^\dagger\right)\,U\right)_{li}\,\chi_{iL} +
  (RV)_{lj}\, \hat{N}_{jL}\,,~l=e,\mu,\tau\,,
  \label{eq:nulL}
\end{equation}
%
where we have kept terms not smaller than $O(R^2)$. It follows from the preceding
relation that in the case under discussion the PMNS neutrino mixing matrix
is given by:
\begin{equation}
 U_\PMNS \;=\; (\I+\eta) U \,,
 \label{PMNS}
\end{equation}
%
where $\eta\;=\;-\half R R^\dagger = -\,\half (RV)(RV)^\dagger$, and $U$, as we have noticed, diagonalises the Majorana mass matrix $m_{\nu}$:
\begin{equation}
 U^\dagger_{\nu}\, m_\nu\, U^{\ast}\;=\;\diag(m_{1},m_{2},m_{3})
\label{U}
\end{equation}
%
The matrix $\eta$ parametrises the deviation from unitarity of the neutrino mixing matrix (\ref{PMNS}). 

The charged current (CC) and the neutral current (NC) weak interaction couplings involving the light Majorana neutrinos $\chi_i$ with definite mass $m_i$
have the form:
\begin{eqnarray}
\label{nuCC}
\mathcal{L}_{CC}^\nu 
&=& -\, \frac{g}{\sqrt{2}}\, 
\bar{\ell}\,\gamma_{\alpha}\,\nu_{\ell L}\,W^{\alpha}\;
+\; {\rm h.c.}
= -\,\frac{g}{\sqrt{2}}\, 
\bar{\ell}\,\gamma_{\alpha}\,
\left( (1+\eta)U \right)_{\ell i}\,\chi_{i L}\,W^{\alpha}\;
+\; {\rm h.c.}\,,\\
\label{nuNC} 
\mathcal{L}_{NC}^\nu &=& -\, \frac{g}{2 \cos\theta_{W}}\,
\overline{\nu_{\ell L}}\,\gamma_{\alpha}\,
\nu_{\ell L}\,
Z^{\alpha}\;  
= -\,\frac{g}{2 \cos\theta_{W}}\,
\overline{\chi_{i L}}\,\gamma_{\alpha}\,
\left (U^\dagger(1+\eta+\eta^\dagger)U\right)_{ik}\,\chi_{k L}\,
Z^{\alpha}\;,
\end{eqnarray}
%
$\theta_{W}$ being the weak mixing angle. It follows from Eq. (\ref{eq:nulL}) that the matrix $RV$ determines the charged current (neutral current) weak interaction couplings of the heavy Majorana neutrinos $\hat{N}_j$ with the charged leptons and the SM $W^\pm$-bosons (flavour neutrinos and the $Z^0$-boson):
\begin{eqnarray}
 \mathcal{L}_{CC}^N &=& -\,\frac{g}{2\sqrt{2}}\, 
\bar{\ell}\,\gamma_{\alpha}\,(1 - \gamma_5)\,(RV)_{\ell j}\,\hat{N}_{j}\,W^{\alpha}\;
+\; {\rm h.c.}\,\label{eq:NCC},\\
 \mathcal{L}_{NC}^N &=& -\frac{g}{4 \cos\theta_{W}}\,
 \overline{\nu_{\ell L}}\,\gamma_{\alpha}\, (1 - \gamma_5)\,(RV)_{\ell j}\,
 \hat{N}_{j}\,Z^{\alpha}\;
+\; {\rm h.c.}\,\label{eq:NNC}.
\end{eqnarray} 
%
Independently of its origin, the mixing of the heavy Majorana neutrinos with the LH flavour neutrinos, parametrised by the matrix $RV$, is constrained via the limits on $\eta = -\,0.5(RV)(RV)^\dagger$ by low energy data. More specifically, electroweak precision data and data on flavour observables set upper bounds on the elements of $\eta$, which in the considered seesaw scenario range
between $9.4\times 10^{-6}$ and $1.3\times 10^{-4}$ (at 95\% C.L.)
\cite{Blennow:2023mqx}. Given such stringent upper bounds on $\eta$,
we have $U_\PMNS \cong  U$ to a very good approximation.

We will assume further that $\hat{N}_j$ have masses not smaller than $\sim 10$ GeV (we will justify this choice later), and that $M_R$ is ``much bigger'' than $M_D$ (at the level of singular values). Using Eq. (\ref{R0}) and the expression for the seesaw neutrino mass matrix (\ref{seesaw}), we obtain at leading order in $R$ the well known relations:
\begin{eqnarray}
  M_R & = &  V\,\hat{M_N}\, V^T (\I + O(R^2))\,,
  \label{M2}
  \\
  R & \simeq & M_D\,M_R^{-1}  \,, 
\label{R}
  \\
m_{\nu} & = & U\hat{m}U^T \; \cong \; -\,R\, M_R\, R^T \,. 
\label{eq:m2} 
\end{eqnarray}
%
Equation (\ref{R}) implies that under the assumptions made the matrix $R$ is indeed ``small''. The usual type I seesaw expression for the Majorana mass matrix of the LH flavour neutrinos is obtained from Eqs. (\ref{R}) and (\ref{eq:m2}): $m_{\nu}\simeq -\,M_{D} M_{R} ^{-1}M_{D}^{T}$.

We note that one can always make $m_i > 0$
\footnote{The lightest neutrino mass can be zero at tree level, but even in
this case receives nonzero higher-order (one- or two-loop) corrections.
Possible sign or phase factors in $m_i$ can be ``absorbed'' by the Majorana
phases in $U_{PMNS}$.}. If $M_R$ is CP-conserving, its eigenvalues $\hat{M_j}$
can be real positive or negative (see, e.g., \cite{Bilenky:1987ty}):
$\hat{M_j} = \rho_j\, M_j$, $\rho_j = \pm 1$, $M_j > 0$. In view of our further
discussion it proves convenient to keep $\rho_j$ in the expression for
$\hat{M_j}$.

In the basis we choose to work and up to corrections $\propto  R R^\dagger$,
the elements of the light neutrino mass matrix $m_{\nu}$ are given by:
$m_{\nu} \equiv U\,\hat{m}\,U^T  \cong  U_{PMNS}\,\hat{m}\,U_{PMNS}^T$. Using the existing upper limits on the absolute scale of neutrino masses and the data on the neutrino mixing angles, obtained in neutrino oscillation experiments, one can obtain the ranges of possible values of the elements of $m_{\nu}$ (see, e.g., \cite{Merle:2006du}). For the purpose of the present study it is sufficient to use the approximate conservative upper bounds $|(m_{\nu})_{l'l}| \lesssim 0.12$ eV, $l,l'=e,\mu,\tau$, which corresponds to the most stringent conservative upper limit on $|(m_\nu)_{ee}|$, obtained in the KamLAND-ZEN neutrinoless double beta decay experiment \cite{KamLAND-Zen:2022tow}.  
From Eq. (\ref{eq:m2}) we obtain to leading order in $R$:
\begin{equation}
|\sum_{j} (RV)_{l'j}\;\rho_j\,M_j\, (RV)^T_{j l}| < 0.12~{\rm eV}\,,
~l',l=e,\mu,\tau\,,~{\rm M_j > 0}\,,~\rho_j = \pm 1\,.
\label{RV1}
\end{equation}
%
Without additional input and barring ``accidental'' cancellations or extreme
fine-tuning (at the level, e.g., of $\sim 10^{11}$ for
$|\hat{M}_j| = M_j \sim 10$ GeV), these constraints can be satisfied by
imposing $| (RV)_{l j}| \sim 10^{-6} \ll 1$ for all $l$ and $j$, which renders
the observation of the heavy Majorana neutrinos $\hat{N}_j$ very difficult,
if not impossible, at any of the experiments aimed at detecting them
(LHC, SHiP, the future planned $e^{+} - e^{-}$ ($FCC_{ee}$, CEPC, ILC),
$\mu^- - \mu^+$ colliders, etc.).

We are interested in the case of sufficiently large $\hat{N}_j$ weak
interaction couplings $|(RV)_{lj}|$ that would permit $\hat{N}_j$ observation.
It is possible to avoid the indicated suppression of $|(RV)_{lj}|$ if,
in particular, there is a strong compensation between the contributions of
the different heavy Majorana neutrinos in the sum in Eq. (\ref{RV1}).
This is most naturally realised in the so-called ``symmetry-protected'' seesaw
scenarios in which there exists an approximately conserved
``non-standard'' lepton charge (see, e.g., \cite{Shaposhnikov:2006nn,Ibarra:2010xw,Gavela:2009cd,Penedo:2017knr,Granelli:2025lds})
\footnote{The ``non-standard'' lepton charges were introduced and the
implications of their approximate conservation were discussed
in \cite{Wolfenstein:1981kw,Petcov:1982ya,Wyler:1982dd,Leung:1983ti,Bilenky:1987ty,Shaposhnikov:2006nn}.} under which both the LH flavour and the RH neutrino fields $\nu_{lL}$
and $\nu_{aR}$  are charged.
In what follows we consider two realistic 
symmetry-protected seesaw scenarios with two RH neutrinos. We also
give an example of a scenario with three RH neutrinos in which
the constraint in Eq. (\ref{RV1}) is satisfied and the heavy
Majorana neutrinos have relatively large CC and NC couplings.
The discussion of these scenarios will set the basis of
our further analysis.

%
\section{Specific  Seesaw Mechanisms}
\label{sec:spseesaw}
%

%
\subsection{Symmetry Protected Seesaw Scenarios with Two RH Neutrinos}
\label{sec:symseesaw}
%

The so-called ``symmetry-protected'' low-scale seesaw scenarios
\cite{Wyler:1982dd,Mohapatra:1986aw,Malinsky:2005bi}
(see also, e.g., \cite{Shaposhnikov:2006nn,Kersten:2007vk,Gavela:2009cd,Ibarra:2010xw,Ibarra:2011xn,Antusch:2015mia,Penedo:2017knr})
rely on a small explicit
breaking of a $U(1)_{L'}$ symmetry, associated with a conserved
non-standard lepton charge $L'$
\cite{Wolfenstein:1981kw,Petcov:1982ya,Leung:1983ti,Bilenky:1987ty}
\footnote{In \cite{Wyler:1982dd} it was shown that the conservation of a
non-standard lepton charge $L'$ by a neutrino Majorana mass term
written in terms of the LH flavour and RH singlet neutrino fields 
is  a sufficient condition to get massless and massive
Dirac neutrino mass eigenstates as a result of its diagonalisation.
The same result was independently derived 
in \cite{Leung:1983ti}, where it was shown as well that the indicated condition 
is also a necessary one, although the authors of \cite{Leung:1983ti}
did not apply these results to some specific seesaw scenario.}.
Some of the versions of seesaw with  two RH neutrinos we consider in the present
subsection are well known.
We discuss them in some detail for completeness.

We recall that in the scenario with two RH neutrinos
one of the three light (Majorana) neutrinos
is massless (at tree level) and hence the neutrino mass spectrum is
hierarchical. Two possible types of hierarchical spectrum are allowed by
the current neutrino data \cite{ParticleDataGroup:2024cfk}:  
$i)$ normal hierarchical (NH), $m_{1}=0$, $m_{2}=\sqrt{\dmsol}$ and
$m_{3}=\sqrt{\dma}$, where $\dmsol \equiv m^2_{2} -  m^2_{1} > 0$ and
$\dma \equiv m^2_{3} -  m^2_{1}$;
$ii)$ inverted hierarchical (IH), $m_{3}=0$, $m_{2}=\sqrt{|\dma|}$ and $m_{1}=\sqrt{|\dma| -\dmsol} \cong \sqrt{|\dma|}$, where $\dma =  m^2_{3} -  m^2_{2} < 0$. In both cases we have: $\dmsol/|\dma| \cong 0.03 \ll 1$.

In what concerns the non-standard lepton charge $L'$,
consider for definiteness the following $L'$ assignments:
$L'(\nu_{lL}) = +\,1$, $l=e,\mu,\tau$, $L'(\nu_{1R}) = +\,1$,
$L'(\nu_{2R}) = -\,1$ 
\footnote{The ``non-standard'' lepton charge is expressed in
terms of the individual lepton charges $L_l$ ($l=e,\mu,\tau$) and
$L_a(\nu_{bR}) = \delta_{ab}$ as $L' = L_e + L_\mu + L_\tau + L_1 - L_2$.
If $L'$ is conserved, then ${\rm min}(n_{+},n_{-})$ and $|n_{+} - n_{-}|$ are
the numbers of massive
Dirac and massless neutrinos, respectively, $n_{+}$ $(n_{-})$ being the number
of charges entering into the expression for $L'$ with positive (negative)
sign \cite{Leung:1983ti} (see also \cite{Bilenky:1987ty}).}.
In the limit of conserved $L'$ the neutrino mass Lagrangian has the form:
\begin{equation}
  \mathcal{L}^{\rm sym}_{\nu}\;=
  \; -\, \overline{\nu_{\ell L}}\,(M_{D})_{\ell 1}\, \nu_{1R} - 
  \half\, (M_{R})_{12}\,\left(\overline{\nu^{C}_{1L}}\,\nu_{2R}\; +\;
\overline{\nu^{C}_{2L}}\,\nu_{1R}\right)\;
  +\;{\rm h.c.}\,, 
\label{eq:Lsym}
\end{equation}
%
Thus, the neutrino Dirac and the Majorana mass matrices
$M_D$ and $M_R$ are given by:
\begin{equation}
	M_D = \left(
     \begin{array}{cc} 
  m^{D}_{e1} & 0 \\
   m^{D}_{\mu 1} & 0 \\
   m^{D}_{\tau 1} & 0
\end{array}\right)\,,\;\;\;\;\;\;\;\;
M_R = \left(
     \begin{array}{cc} 
     0 & M_{12}\\
    M_{12} & 0
\end{array}\right)\,.
\label{eq:MDMR1}
\end{equation}
%
We can take $(M_{R})_{12}$ to be real and positive without loss of generality,
$((M_{R})_{12})^* =  (M_{R})_{12} > 0$, 
and we assume for simplicity that $m^{D}_{l1}$, $l=e,\mu,\tau$, are real.
The diagonalisation of the neutrino mass matrix in Eq. (\ref{seesaw})
reveals that there are three massless neutrinos, $n_{1,2,3}$, and two
heavy Majorana neutrino mass eigenstates $N_{1,2}$ with equal masses,
$M_1 = M_2 = \sqrt{M^2_{12} + \sum_l  (m^{D}_{l1})^2} \cong M_{12}$:
\begin{equation}
  N_1 = \gamma_5\,\hat{N_1} = \dfrac{1}{\sqrt{2}}\left(\nu_{1R} - \nu^{C}_{1L} - \nu_{2R}
  + \nu^{C}_{2L}\right)\,,~~
  N_2 = \hat{N_2} = \dfrac{1}{\sqrt{2}}\left(\nu_{1R} + \nu^{C}_{1L} + \nu_{2R}
  + \nu^{C}_{2L}\right)\,.
  \label{eq:N12sym}
\end{equation}
%
The fields of $N_{1,2}$  satisfying the Majorana conditions
\footnote{Since 
$\hat{N}_{1} = \gamma_5\,N_1$, where $\hat{N}_{1}$ has a negative mass
$\hat{M}_1 = -\, M_1 < 0$, 
the terms involving $\hat{N}_{1}$ in Eqs. (\ref{eq:NCC}) and (\ref{eq:NNC})
change sign when written in terms of $N_1$ of interest:
$(1 - \gamma_5)\,(RV)_{\ell 1}\,\hat{N}_{1} =
-\,(1 - \gamma_5)\,(RV)_{\ell 1}\,N_{1}$.}:
\begin{equation}
  C\, (\overline{N_j})^T = \rho_j\,N_j\,
    \quad \rho_1 = -\,1\,,~~\rho_2 = +\,1\,.
 \label{eq:MajcondN12}
\end{equation}
%
The RH neutrino Majorana mass matrix given in 
Eq. (\ref{eq:MDMR1}) is diagonalised by the orthogonal matrix
$V(\varphi)$ with $\varphi = \pi/4$:
\begin{equation}
V = \left(
     \begin{array}{cc} 
    \cos\varphi & \sin\varphi\\
    -\,\sin\varphi & \cos\varphi
     \end{array}\right) =
  \dfrac{1}{\sqrt{2}}\left(
     \begin{array}{cc} 
    1 & 1 \\
    -\,1 & 1
     \end{array}\right)
 \label{eq:Vsym}
\end{equation}
%

The RH neutrino Majorana mass term in Eq. (\ref{eq:Lsym})
conserves the lepton charge $L_1 - L_2$.
The fields $N_{1,2}$ form a Dirac field:
$N_D = (N_1 + N_2)/\sqrt{2}$. They couple to the weak $W^\pm$ and $Z^0$ bosons
only through the combination $(N_{1L} + N_{2L})/\sqrt{2}$, i.e.,
only through the LH component of the Dirac field $N_D$:
$(RV)_{l2} = -\,(RV)_{l1} = m^{D}_{l1}/(\sqrt{2}M_{12})$,
$-\,(RV)_{l1}N_{1L} +(RV)_{l2}N_{2L} =
(m^{D}_{l1}/M_{12})(N_{1L} + N_{2L})/\sqrt{2}$.
The light neutrino Majorana mass matrix
$m_\nu = -\,M_D\,M^{-1}_R\,M_D^T = 0$.

All these properties follow from the conservation of the non-standard
lepton charge $L^\prime = L_e + L_\mu + L_\tau + L_1 - L_2$, of course,
and are well known. The massless neutrinos $n_i$, $i=1,2,3$, are the
dominant components of the LH flavour neutrino fields $\nu_{lL}$,
while $N_{DR}$ and $N^C_{DR}$ are the dominant components of the
two RH neutrino fields $\nu_{1R}$ and $\nu_{2R}$.

  It should be clear that in the approximation being discussed,
there are no physical (observable) effects associated with the fact that 
$N_{1,2}$ are Majorana particles: $N_{1,2}$ always appear in the interaction 
Lagrangian in the combination $(N_{1L} + N_{2L})/\sqrt{2}$ which 
is equivalent to a Dirac fermion.  

\vspace{-0.4cm}
\subsubsection{ Breaking of $L'$-symmetry by the neutrino Yukawa coupling.}
\label{sec:LpB1}

\vspace{-0.4cm}
The considered $L'$-symmetry has to be broken to generate non-zero
neutrino masses. The breaking  can be generated by the neutrino
Yukawa coupling and/or by diagonal elements of the Majorana mass matrix
of the RH neutrinos, which lead respectively to the following additional
terms in $\mathcal{L}^{\rm sym}_{\nu}$:
\begin{equation}
  \mathcal{L}^{\rm SB}_{\nu D}\;=
  \; -\, \overline{\nu_{\ell L}}\,(m_{D})_{\ell 2}\, \nu_{2R}\;+\;{\rm h.c.}\,, 
\label{LSBY}
\end{equation}
and
\begin{equation}
  \mathcal{L}^{\rm SB}_{\nu M}\;=\; -\, 
  \half\, \left ( \overline{\nu^{C}_{1L}}\,(M_{R})_{11}\,\nu_{1R} +
  \overline{\nu^{C}_{2L}}\,(M_{R})_{22}\,\nu_{2R} \right) \;+\;{\rm h.c.}\,. 
\label{LSBM}
\end{equation}
%
Consider first the case of breaking due only to the neutrino
Yukawa coupling: $(m_{D})_{\ell 2}\neq 0$,
$(M_{R})_{11} = (M_{R})_{22} = 0$
\footnote{This is usually called the ``linear'' seesaw mechanism
\cite{Malinsky:2005bi}.}.
We take again $(m_{D})_{\ell 2}$ to be real
for simplicity. The neutrino Dirac mass matrix now has the form:
\begin{equation}
	M_D = \left(
     \begin{array}{cc} 
  m^{D}_{e1} &  m^{D}_{e2} \\
   m^{D}_{\mu 1} &  m^{D}_{\mu 2}  \\
   m^{D}_{\tau 1} &  m^{D}_{\tau 2}
\end{array}\right)\,.
\label{eq:MD12}
\end{equation}
%
The RH neutrino Majorana mass matrix is the same as in Eq. (\ref{eq:MDMR1}).
In this case (see, e.g., \cite{Ibarra:2010xw,Penedo:2017knr}) 
i) $(m_\nu)_{ll'} = -\, (m^{D}_{l1} m^{D}_{l'2}+m^{D}_{l2} m^{D}_{l'1})/M_{12}\neq 0$,
ii) the heavy Majorana neutrinos have different masses, $M_2 \neq M_1$,
and now form a pseudo-Dirac pair \cite{Wolfenstein:1981kw,Petcov:1982ya},
$N_{PD} = (N_{1} + N_{2})/\sqrt{2}$.
iii) the angle $\varphi = \pi/4$ in the $V$ matrix gets a negligible
correction $\sim |(m^{D}_{e1} m^{D}_{e2} + m^{D}_{\mu 1} m^{D}_{\mu2}
+ m^{D}_{\tau 1} m^{D}_{\tau 2})|/M^2_{12} \ll 1$.
As a consequence of iii), in particular,  we have:
$(RV)_{l1} = (m^{D}_{l2} -  m^{D}_{l1})/(\sqrt{2}M_{12})$,
$(RV)_{l2} = (m^{D}_{l2} +  m^{D}_{l1})/(\sqrt{2}M_{12})$.
The $N_{1,2}$ mass splitting is determined by the light neutrino
masses \cite{Shaposhnikov:2006nn,Ibarra:2010xw,Penedo:2017knr}.
In the case of NH neutrino mass ordering,
for example, $M_2 = M_{12} + 0.5( m_3 - m_2)$, $M_1 = M_{12} - 0.5( m_3 - m_2)$,
$(M_2 - M_1) = m_3 - m_2 \cong 0.0414$ eV
\footnote{We have used $m_3 \cong 5.01\times 10^{-2}$ eV and
$m_2 \cong 0.865\times 10^{-2}$ eV \cite{Esteban:2024eli}.
In the IH case $(M_2 - M_1)$ is even smaller.},
and, thus, it is exceedingly small and hardly directly observable in practice.

From the constraint  $|(m_\nu)_{ll'}| < 0.12$ eV discussed earlier,
we get for $M_{1,2} \cong M_{12} = 10$ GeV:
$|(m^{D}_{l1}m^{D}_{l'2}+m^{D}_{l2}m^{D}_{l'1})|<10^{-9}~{\rm GeV^2~(M_{12}/10 GeV)}$.
The only possibility to have a relatively large $N_{1,2}$ CC and NC couplings
and thus possibly observable $N_{1,2}$ production rates,
is to assume that, e.g., $|m^{D}_{l2}| \ll |m^{D}_{l'1}|$. This possibility was
investigated in detail in \cite{Penedo:2017knr}, 
where also a theoretical justification for the inequality of interest was
presented. Taking, for example, $|m^{D}_{l2}|\sim 10^{-8}$ GeV and
$|m^{D}_{l1}|\lesssim 0.10$ GeV
\footnote{Actually, values of  $|m^{D}_{l2}|\sim 10^{-8}$ GeV
are required to reproduce, e.g., the values of the two neutrino mass-squared
differences and of the neutrino mixing angles measured in neutrino
oscillation experiments.},
we have $|(RV)_{l1}| \cong |(RV)_{l2}|\lesssim
|m^{D}_{l1}/(\sqrt{2}M_{12})| \lesssim 7\times 10^{-3}\,(10~{\rm GeV}/M_{12})$.
Thus, in principle, $N_1$ and $N_2$ can be 
produced with sufficiently large rates at, e.g., $e^+ - e^-$ colliders, 
which might allow them to be observed.
Since $(RV)_{\ell 2} = -\,(RV)_{\ell 1} \cong m^{D}_{\ell 1}/(\sqrt{2}M_{12})$, 
$N_1$ and $N_2$ couple 
to a given charged lepton $l^-$ (flavour neutrino $\nu_l$) 
in the weak charged (neutral) current always in the 
combination $(N_{1} + N_{2})/\sqrt{2} = N_{PD}$.
In the limit of $M_1 = M_2$,  $N_{PD}$ behaves as a Dirac particle. 
Therefore the magnitude of 
all physical effects associated with the Majorana nature of 
$N_1$ and $N_2$ is  determined by 
the mass difference $M_2 - M_1$.
The fact that  $M_2 - M_1$ is extremely small,
being proportional to the light-neutrino mass difference,
seems to indicate that the Majorana  effects related to $N_{1,2}$ would be 
difficult to observe experimentally.

We note finally that, as was shown in \cite{Sandner:2023tcg},
successful leptogenesis is possible in the considered seesaw scenario
for couplings and values of $M_{1,2}$ that can be probed at FCCee during its
first phase of operation running at the $Z^0$-pole.

%
\subsubsection{Breaking of  $L^\prime$-symmetry by the neutrino Yukawa coupling and the
  RH neutrino Majorana mass term.}
\label{sec:LpB3}
%

In this case the neutrino Dirac and the RH Majorana mass matrices have
the form:
\begin{equation}
	M_D = \left(
     \begin{array}{cc} 
  m^{D}_{e1} & m^{D}_{e2} \\
   m^{D}_{\mu 1} & m^{D}_{\mu 2}  \\
   m^{D}_{\tau 1} & m^{D}_{\tau 2}
\end{array}\right)\,,\;\;\;\;\;\;\;\;
M_R = \left(
     \begin{array}{cc} 
     0 & M_{12}\\
    M_{12} & M_{22}
\end{array}\right)\,,
\label{eq:MDMRLB2}
\end{equation}
%
with the full neutrino mass matrix 
\begin{equation}
\mathcal{M} =  \left(\begin{array}{cc}
	\N & M_D \\
	M_D^T & M_R \\        
       \end{array}
 \right)\,.
 \label{eq:MnuLpBB}
\end{equation}
%
Here $M_{22}\neq 0$, $|M_{22}| < (\ll) M_{12}$ and for 
concreteness $M_{22}$ is taken to be real and positive,
while $m^{D}_{l1}$ and $m^{D}_{l'2}$, $l,l'=e,\mu,\tau$, 
are taken to be real
\footnote{Setting $m^{D}_{l2} = 0$, $l=e,\mu,\tau$,
corresponds to the inverse seesaw mechanism \cite{Wyler:1982dd,Mohapatra:1986aw}.
In the seesaw scenario
with $m^{D}_{l2} = 0$, 
the light neutrino Majorana mass matrix $m_\nu$ has two zero eigenvalues
at tree level
and the second nonzero light neutrino mass and the neutrino
mixing are generated radiatively \cite{Adhikari:2010yt,Dev:2012sg}.
In what concerns
the test of the Majorana nature of the two heavy Majorana neutrinos,
the scenario investigated in \cite{Adhikari:2010yt,Dev:2012sg} is a limiting
case of the one considered in the present subsection.
}.
In this scenario the flavour neutrino Majorana mass matrix is
given by:
\begin{equation}
  (m_\nu)_{ll'} = -\, \dfrac{1}{M_{12}}
  \left (m^{D}_{l1}m^{D}_{l'2} + m^{D}_{l2}m^{D}_{l'1}
  -\,\dfrac{M_{22}}{M_{12}}m^{D}_{l1}m^{D}_{l'1} \right)\,,~~l,l'=e,\mu,\tau\,.
\label{eq:mnuLpBB}
\end{equation}
%
The eigenvalues of $M_R$ are
$M_{1,2} =0.5(\sqrt{4M^2_{12} + M^2_{22}} \mp M_{22})$, with $\rho_1 = -1$,
$\rho_2 = 1$. So, we have:  $M_2 - M_1 = M_{22}$ and $M_1M_2 = M^2_{12}$.
The angle $\varphi$ of the $V$-matrix can be expressed in terms of $M_{1,2}$:
$\cos^2\varphi = M_2/(M_2 + M_1)$, $\sin^2\varphi = M_1/(M_2 + M_1)$.

The $RV$ couplings read:
\begin{equation}
  (RV)_{l1} = \dfrac{ m^{D}_{l2}\sin\varphi - m^{D}_{l1}\cos\varphi}{M_{1}}\,,
  \quad
  (RV)_{l2} = \dfrac{ m^{D}_{l2}\cos\varphi + m^{D}_{l1}\sin\varphi}{M_{2}}\,,~
  l=e,\mu,\tau\,.
     \label{eq:RVl1l2}
\end{equation}
%
For the inequality given in Eq. (\ref{RV1}) we get:
\begin{equation}
  \left |-\,(RV)_{l1}M_1(RV)_{l'1} + (RV)_{l2}M_2(RV)_{l'2}\right |
  = \left |  m^{D}_{l1} m^{D}_{l'1}\dfrac{M_2 - M_1}{M_1M_2}
  -\,\dfrac{m^{D}_{l1} m^{D}_{l'2}
    +  m^{D}_{l2}m^{D}_{l'1}}{\sqrt{M_1M_2}}\right| < 0.12~{\rm eV}\,.
\label{eq:RMRR}
\end{equation}
%
We are interested in the case of i) relatively large heavy Majorana neutrino
$|(RV)_{lj}|$-couplings and ii) sizeable mass splitting $(M_2 - M_1)$.
Under these requirements and barring accidental cancellations, the inequality in Eq. (\ref{eq:RMRR}) can be satisfied
only  if  $|m^{D}_{l1}| \ll  |m^{D}_{l'2}|$, $l,l'=e,\mu,\tau$.
Taking as a reference value $\sqrt{M_1M_2} = M_{12} = 10$ GeV,
the inequality (\ref{eq:RMRR}) does not constrain $M_2 - M_1$ and leads
to a mild constraint on  $|m^{D}_{l2}|$ for, e.g., $|m^{D}_{l1}|\sim 10^{-9}$ GeV.
Indeed, for $|m^{D}_{l1}| = 10^{-9}$ GeV we get approximately 
$|m^{D}_{l2}|< 0.6~{\rm GeV}~(M_{12}/10~{\rm GeV})$.
In the considered case the terms
$\propto m^{D}_{l1}$ in the expressions for $(RV)_{l1}$ and $(RV)_{l2}$ in
Eq. (\ref{eq:RVl1l2}) can be neglected.

The  relative mass difference $(M_2 - M_1)/M_{12}$ and $m^{D}_{e2}/M_{12}$ are
constrained by the combined limits on the neutrinoless double beta
($\beta\beta_{0\nu}$) decay
effective Majorana mass $|<m>|$ and on the parameter
$\eta_{ll'} = -\,0.5(RV)_{lj}(RV)^\dagger_{jl'}$ characterising the deviation
from unitarity of the PMNS matrix.
For $|m^{D}_{l1}|\lesssim 10^{-9}$ GeV and $M_{12} \ge 10$ GeV of interest, 
the heavy Majorana neutrino contribution to  $|<m>|$ has the form:
\begin{equation}
  |<m>^N| \cong f(A)\dfrac{M^2_a}{M_{12}}\,
  \left(\dfrac{m^{D}_{e2}}{M_{12}}\right)^2\,\dfrac{M_2-M_1}{M_{12}}\,,
  \label{eq:meffN}
\end{equation}
%
where $f(A)$ is the nuclear matrix element (NME) of the decay,
$A$ being the mass number of the decaying nucleus, 
and $M_a \cong 0.9$ GeV (see, e.g., \cite{Blennow:2010th}).
The function $f(A)$ exhibits weak dependence
on $A$: for $^{76}$Ge, $^{82}$Se, $^{130}$Te and $^{136}$Xe it takes the values
\cite{Blennow:2010th}
of 0.079,  0.073,  0.080 and  0.068, respectively.
The heavy Majorana neutrino contribution also should satisfy 
$|<m>^N| < 0.12$ eV.
From Eq. (\ref{eq:meffN}) we get:
\begin{equation}
  |M_2-M_1| < 0.12~{\rm eV}\, \dfrac{M^2_{12}}{f(A)M^2_a}\,\left(\dfrac{M_{12}}{m^D_{e2}}\right)^2\,.
  \label{eq:M2minM1}
\end{equation}
%
Thus, in order to have a sufficiently large and possibly observable
mass difference $(M_2-M_1)$, which is crucial for testing and revealing
the Majorana nature of $N_{1,2}$,
$M_{12}$ has to be sufficiently large. For this reason we concentrate on the
case of $M_{12} \gtap 10$ GeV.
At the same time, the
parameter $\eta_{ll'}$ is given by:
\begin{equation}
  \eta_{ll'} = -\,\dfrac{1}{2}\,\dfrac{m^{D}_{l2}\,m^{D}_{l'2}}{M^2_{12}}\,.
  \label{eq:etallp}
  \end{equation}
%
The limits on $|\eta_{ll'}|$ \cite{Blennow:2023mqx}
imply (at 95\% C.L.):
\begin{align}
\begin{aligned}
&\left |\dfrac{m^{D}_{e2}}{M_{12}}\right|^2 < 1.9\times 10^{-5}~(1.10\times 10^{-3})\,,~
  \left |\dfrac{m^{D}_{\mu2}}{M_{12}}\right|^2 < 2.60\times 10^{-4}~(6.4\times 10^{-5})\,,~ 
\dfrac{|m^{D}_{\mu 2}m^{D}_{e2}|}{M^2_{12}} < 2.40\times 10^{-5}~(2.60\times 10^{-5})\,,\\[0.25cm]
&\left |\dfrac{m^{D}_{\tau 2}}{M_{12}}\right|^2 < 4.20\times 10^{-4}(9.0\times 10^{-5})\,,~
\dfrac{|m^{D}_{\tau 2}m^{D}_{e2}|}{M^2_{12}} < 4.4\times 10^{-5}(2.80\times 10^{-4})\,,~
\dfrac{|m^{D}_{\tau 2}m^{D}_{\mu 2}|}{M^2_{12}} < 2.60\times 10^{-4}(7.0\times 10^{-5})\,,
\label{eq:etalimits}
\end{aligned}
\end{align}
%
where the quoted values (values in brackets) correspond to the NH (IH) light-
neutrino mass spectrum.
As we will see, the ratio $(m^{D}_{l2}/M_{12})^2$ determines the rate of
$N_{1,2}$ production in $e^+ - e^-$ collisions. Therefore it should be
sufficiently large for $N_{1,2}$ to be produced with observable rates.

All the listed constraints and requirements can be satisfied, e.g.,
for $m^{D}_{l2} \lesssim 3.3\times 10^{-3}$ GeV and $M_{12}\gtap 10$ GeV.
For these values of $m^{D}_{l2}$ and  $M_{12}$, the constraints on
the mass difference of interest are rather weak:
$|M_2 - M_1| < 2.1~(M_{12}/(10~{\rm GeV}))^4$ GeV, where we have used
$f(A) = 0.068$. However, we  always require that
$(M_2 - M_1) < (\ll)\,M_{12}$.

For relatively large values of $m^{D}_{l2}$ and the values of $M_{1,2}$
of interest, there might be significant one-loop corrections to
$(m_\nu)_{ll'}$. These corrections are given by
\cite{Pilaftsis:1991ug,Grimus:2002nk,AristizabalSierra:2011mn,Dev:2012sg,
 Lopez-Pavon:2012yda}
(see also \cite{Lopez-Pavon:2015cga}):
\begin{equation}
( m^{\text{1-loop}}_\nu)_{ll'} =
\sum_{j=1}^2 (RV)_{lj} \left( \frac{\rho_jM^3_j}{16 \pi^2 v^2}
\left(\frac{\ln\left(\frac{M^2_j}{M_H^2}\right)}
     {\frac{M^2_j}{M_H^2}-1} +
3 \frac{\ln\left(\frac{M^2_j}{M_Z^2}\right)}{\frac{M^2_j}{M_Z^2}-1}\right)
\right) (RV)_{l'j}\,,
  \label{eq:1loopmnu}
\end{equation}
%
where $M_H$ is the SM Higgs boson mass,
$M_H \cong 125$ GeV, $M_Z \cong 91$ GeV, and $v\cong 246$ GeV is the
Higgs vacuum expectation value. 
Thus, the total light neutrino Majorana mass matrix is
$m^{{\rm total}}_{\nu} = m_\nu + m^{\text{1-loop}}_\nu$. 
We note that $m^{\text{1-loop}}_\nu$ vanishes in the case of  $M_2 = M_1$.
Taking into account that to a good approximation
$(RV)_{l1} \cong m^{D}_{l2}/\sqrt{M_1(M_1+M_2)}$ and
$(RV)_{l2} \cong m^{D}_{l2}/\sqrt{M_2(M_1+M_2)}$ we find that 
in order to keep the elements of $ m^{\text{1-loop}}_\nu$
sufficiently small for the values of $M_{1,2} \gtap 10$ GeV
and $(M_2 -M_1) \sim (1- 4)$ GeV considered by us (see further),
so that $|(m^{\text{1-loop}}_\nu)_{ll'}| \ll 10^{-2}$ eV, 
it is sufficient to have $m^{D}_{l2} \le 10^{-3}$ GeV.
In the analysis which follows we will use the benchmark values 
$m^{D}_{l2} =  10^{-3}$ GeV, $l=e,\mu,\tau$
\footnote{
The contribution from two-loop corrections is  small as it 
is suppressed by an extra factor $\propto m^{D}_{l2} m^{D}_{l'2}$ and a further
factor $\mathcal{O}(10^{-2})$ from the loop integral.}.
We always take  $M_{22} \lesssim 0.1\, M_{12}$ as well.

We note finally that the diagonalisation of the full neutrino mass matrix
$\mathcal{M}$ given in 
Eq. (\ref{eq:MnuLpBB}) leads to corrections of $O((m^{D}_{l1}/M_{12})^2)$ and 
$O((m^{D}_{l2}/M_{12})^2)$ to both the masses $M_{1,2}$ and the angle
$\varphi$ of the matrix $V$ discussed so far. For the values of
$|m^{D}_{l1}|$ and $M_{12}$ we consider, the corrections of
$O((m^{D}_{l1}/M_{12})^2)$ are totally negligible. Taking into account
the corrections of $O((m^{D}_{l2}/M_{12})^2)$ 
modifies $M_{1,2}$ to
$M'_{1,2} = 0.5[\sqrt{M^2_{22} + 4(M^2_{12} + \sum_l(m^{D}_{l2})^2)} \mp M_{22}]$,
  so still $M'_{2} - M'_{1} = M_{2} - M_{1}$. Given the constraints
  (\ref{eq:etalimits}) on $(m^{D}_{l2})^2/M^2_{12}$, the corrections
  $\sim (m^{D}_{l2})^2/M^2_{12}$ in $M'_{1,2}$ can be safely neglected and
  to a good approximation $M'_{1,2} = M_{1,2}$.
  One can  show that the discussed correction to the value of
  the angle $\varphi$ is also negligibly small.

  There is an alternative possibility to
  get a relatively large $N_1 - N_2$ mass splitting. It consists of 
  adding one more $L'$-symmetry breaking parameter, namely $M_{11}$
  (see Eq. (\ref{LSBM}))
  in the RH neutrino Majorana mass matrix which takes the form:
  \begin{equation}
M_R = \left(
     \begin{array}{cc} 
     M_{11} & M_{12}\\
    M_{12} & M_{22}
\end{array}\right)\,,
\label{eq:MDMRLB3}
\end{equation}
  %
  where, as in the previously discussed model, $0 < M_{22} \ll M_{12}$,
  while we have also, e.g., $|M_{11}| \gg M_{22},m^D_{l2}$,
  $|M_{11}|\gtap  M_{12}$
  \footnote{This version is often referred to as an ``extended'' seesaw model
  \cite{Kang:2006sn} (see also, e.g., \cite{Lopez-Pavon:2012yda})}.
  The phenomenology of the low-scale version of this seesaw model
  was discussed in  detail in \cite{Lopez-Pavon:2012yda}. 
  It was shown, in particular, that $M_{11}$ might have practically
  no effect on the generation of
  light neutrino Majorana mass matrix $m_\nu$  and thus might not be
  constrained by the smallness of the elements of $m_\nu$. At the same time
  $M_{11}$ contributes to the mass difference $(M_2 - M_1)$.
  The parameter $M_{11}$ can be constrained
  using the limits on the $N_{1,2}$ contributions to the
  $(\beta\beta)_{0\nu}$-decay 
  effective Majorana mass and by the requirement of
  sufficiently small one-loop corrections to $m_\nu$.
  Taking into account these constraints still allows a sizeable
  $N_2 - N_1$ mass splitting.
  This case was discussed in detail in
  \cite{Hernandez:2018cgc} with the aim of
  evaluating the prospective sensitivity to the splitting of
  the experiments at the ILC $e^+ - e^-$ collider.
  
  From the discussion in this subsection
  it is clear that the mass splitting $|M_2 - M_1|$, and thus
  the lepton charge violating effects related to the Majorana
  nature of the heavy neutrinos $N_{1,2}$,  can be relatively large
  and possibly observable in some of 
  the symmetry-protected seesaw models with two RH neutrinos.
 
%
\subsection{Seesaw Scenario with Three RH Neutrinos}
\label{sec:3RH}
%
%

It is possible to avoid the suppression of $N_j$ weak interaction
couplings $|(RV)_{lj}|$ due to the constraint given in Eq. (\ref{RV1}) also in the
case when all three heavy Majorana neutrinos contribute in the sum in
Eq. (\ref{RV1}) when there is a strong mutual compensation between their
contributions. We provide one such example of masses and couplings below 
from \cite{Granelli:2026qss}.
It is a particular version of the seesaw scenarios investigated
recently in \cite{Granelli:2026qss}. These scenarios
reproduce successfully the neutrino oscillation data on the two light neutrino
mass-squared differences and the three neutrino mixing angles.
The observed BAU is generated via the leptogenesis mechanism
in which the requisite CP-violation is provided exclusively by the Dirac phase
in the PMNS matrix.

In the example considered here, the three heavy Majorana neutrinos $N_{1,2,3}$ are assumed to have masses
$M_1 = 9$ GeV, $M_3 = 10$ GeV and $M_2 = 11$ GeV ($\rho_{1,2,3} = 1$). So, their
mass splittings are relatively large: we have $\delta M_{13} = 0.105$,
$\delta M_{12} =0.200$, $\delta M_{23} =  0.095$, where
$\delta M_{r r'} \equiv |M_r - M_{r'}|/(0.5(M_r + M_{r'}))$. The matrix of the
charged and neutral current couplings of $N_{1,3,2}$ reads:
\begin{eqnarray}
  \nonumber
& RV = 
\begin{pmatrix}
(RV)_{e 1} & (RV)_{e 3} & (RV)_{e 2}\\
(RV)_{\mu 1} & (RV)_{\mu 3} & (RV)_{\mu 2}\\
 (RV)_{\tau 1} & (RV)_{\tau 3} & (RV)_{\tau 2}   
  \end{pmatrix}
  \\[0.3cm]
  &= \begin{pmatrix}
(-7.878330 -i1.964103)10^{-5} &(-2.721114 +i10.92773)10^{-5} & (+7.600717 +i1.890956)10^{-5}\\
(+21.78472+i0.5495799)10^{-5} & (+0.7625350-i30.03123)10^{-5} &(-20.77582-i0.5305003)10^{-5}\\
(-2.656531+i4.744249)10^{-6} & (+6.582584+i1.963658)10^{-6} & (+0.2918793-i4.579545)10^{-6}
\end{pmatrix}.            
\label{eq:3NsRV}
\end{eqnarray}
%
The couplings of $N_{1,3,2}$ to $e^\pm$ ($\nu_e$ and $\bar{\nu}_e$) and
$\mu^\pm$ ($\nu_\mu$ and $\bar{\nu}_\mu$) are also relatively large:
$|(RV)_{e 1,3,2}|^2 \cong 6.59,12.68,6.13 \times 10^{-9}$,
$|(RV)_{\mu 1,3,2}|^2 \cong 4.749,9.024,4.319 \times 10^{-8}$. At the same time
the couplings to $\tau^\pm$ ($\nu_{\tau}$ and $\bar{\nu}_\tau$) are
significantly smaller, $|(RV)_{\tau 1,3,2}|^2\sim 10^{-11}$.
The constraint in Eq. (\ref{RV1}), as well as the experimental limits on the
$N_{1,2,3}$ contributions to the $(\beta\beta)_{0\nu}$-decay 
effective Majorana mass,
$|<m>^N|$, are satisfied due to a significant mutual
compensation between the three terms in the sum in the left-hand side of the
inequality in Eq. (\ref{RV1}) and in the expression for $|<m>^N|$
\footnote{We have  $|<m>^N| \cong 10^{-3}$ eV.}.
Due to this compensation, the one-loop contribution to
$m_\nu$ is suppressed, but is not negligible \cite{Granelli:2026qss}.

%
\section{Dirac-Majorana Dependent Signatures in Experiments at $e^+ - e^-$ ($\mu^+ - \mu^-$) Colliders}
\label{sec:DvsM}
%
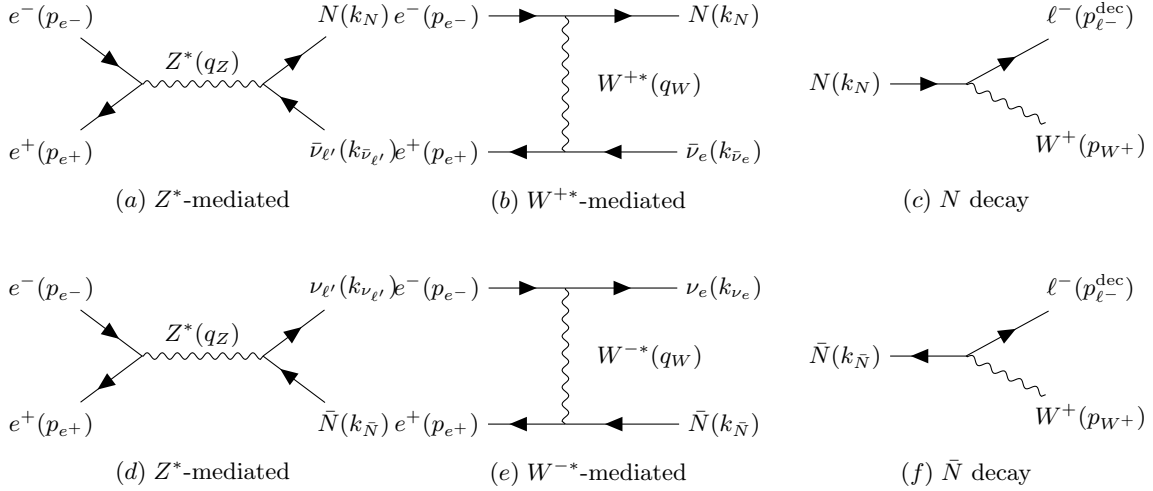
\begin{figure}[t]
\centering

\begin{tikzpicture}

\begin{scope}[shift={(0,3.6)}]
\begin{feynman}

\vertex (em)   at (-2.0,  0.9) {$e^-(p_{e^-})$}; 
\vertex (ep)   at (-2.0, -0.9) {$e^+(p_{e^+})$}; 

\vertex (a1)   at (-0.8, 0);
\vertex (a2)   at ( 0.8, 0);

\vertex (N)    at ( 2.0,  0.9) {$N(k_N)$};
\vertex (nueb) at ( 2.0, -0.9) {$\bar{\nu}_{\ell'}(k_{\bar{\nu}_{\ell'}})$};
\diagram*{
    (em) -- [fermion] (a1),
    (ep) -- [anti fermion] (a1),

    (a1) -- [boson, edge label={$Z^*(q_Z)$}] (a2),

    (a2) -- [fermion] (N),
    (a2) -- [anti fermion] (nueb),
};

\end{feynman}

\node at (0,-1.55) {$(a)\; Z^*$-mediated};

\end{scope}

\begin{scope}[shift={(5.1,3.6)}]
\begin{feynman}

\vertex (em)   at (-2.0,  0.9) {$e^-(p_{e^-})$};
\vertex (ep)   at (-2.0, -0.9) {$e^+(p_{e^+})$}; 

\vertex (b1)   at (-0.3,  0.9);
\vertex (b2)   at (-0.3, -0.9);

\vertex (N)    at ( 1.8,  0.9) {$N(k_N)$};
\vertex (nueb) at ( 1.8, -0.9) {$\bar{\nu}_e(k_{\bar{\nu}_e})$};

\diagram*{
    (em) -- [fermion] (b1)
         -- [fermion] (N),

    (ep) -- [anti fermion] (b2)
         -- [anti fermion] (nueb),

    (b1) -- [boson] (b2),
};

\node[right] at (0.0,0) {$W^{+*}(q_W)$};

\end{feynman}

\node at (0,-1.55) {$(b)\; W^{+*}$-mediated};

\end{scope}

\begin{scope}[shift={(10.1,3.6)}]
\begin{feynman}

\vertex (N)    at (-1.6,0) {$N(k_N)$};
\vertex (c1)   at (0,0);

\vertex (lm)   at (1.65,  0.9)
    {$\ell^-(p_{\ell^-}^{\rm dec})$};

\vertex (Wp)   at (1.65, -0.8)
    {$W^+(p_{W^+})$};

\diagram*{
    (N)  -- [fermion] (c1),
    (c1) -- [fermion] (lm),
    (c1) -- [boson]   (Wp),
};

\end{feynman}

\node at (0,-1.55) {$(c)\; N$ decay};

\end{scope}

\begin{scope}[shift={(0,0)}]
\begin{feynman}

\vertex (em)   at (-2.0,  0.9) {$e^-(p_{e^-})$};
\vertex (ep)   at (-2.0, -0.9) {$e^+(p_{e^+})$};

\vertex (d1)   at (-0.8,0);
\vertex (d2)   at ( 0.8,0);

\vertex (nue)  at ( 2.0,  0.9) {$\nu_{\ell'}(k_{\nu_{\ell'}})$};
\vertex (Nb)   at ( 2.0, -0.9) {$\bar{N}(k_{\bar N})$};

\diagram*{
    (em) -- [fermion] (d1),
    (ep) -- [anti fermion] (d1),

    (d1) -- [boson, edge label={$Z^*(q_Z)$}] (d2),

    (d2) -- [fermion]      (nue),
    (d2) -- [anti fermion] (Nb),
};

\end{feynman}

\node at (0,-1.55) {$(d)\; Z^*$-mediated};

\end{scope}

\begin{scope}[shift={(5.1,0)}]
\begin{feynman}

\vertex (em)   at (-2.0,  0.9) {$e^-(p_{e^-})$};
\vertex (ep)   at (-2.0, -0.9) {$e^+(p_{e^+})$};

\vertex (e1)   at (-0.3,  0.9);
\vertex (e2)   at (-0.3, -0.9);

\vertex (nue)  at ( 1.8,  0.9) {$\nu_e(k_{\nu_e})$};
\vertex (Nb)   at ( 1.8, -0.9) {$\bar{N}(k_{\bar N})$};

\diagram*{
    (em) -- [fermion] (e1)
         -- [fermion] (nue),

    (ep) -- [anti fermion] (e2)
         -- [anti fermion] (Nb),

    (e1) -- [boson] (e2),
};

\node[right] at (0.0,0) {$W^{-*}(q_W)$};

\end{feynman}

\node at (0,-1.55) {$(e)\; W^{-*}$-mediated};

\end{scope}

\begin{scope}[shift={(10.1,0)}]
\begin{feynman}

\vertex (Nb)   at (-1.6,0) {$\bar{N}(k_{\bar N})$};
\vertex (f1)   at (0,0);

\vertex (lm)   at (1.65,  0.9)
    {$\ell^-(p_{\ell^-}^{\rm dec})$};

\vertex (Wp)   at (1.65, -0.8)
    {$W^+(p_{W^+})$};

\diagram*{
    (Nb) -- [anti fermion] (f1),

  (f1) -- [fermion] (lm),

    (f1) -- [boson] (Wp),
};

\end{feynman}

\node at (0,-1.55) {$(f)\; \bar{N}$ decay};

\end{scope}

\end{tikzpicture}

\caption{
Tree-level production and decay diagrams for
$e^- + e^+ \to N+\overline{\nu}_{\ell'}~(\overline{\nu}_e)$
and
$e^- + e^+ \to \bar{N} + \nu_{\ell'}~(\nu_e)$
in the case of real or virtual Majorana $N \equiv \bar{N}$, 
$\ell',\ell = e,\mu,\tau$. For a Dirac $N$ the decay
$\bar{N}\to\ell^- + W^+$ is forbidden. In the case of $\mu^- - \mu^+$
initial beams one has to replace $\nu_e$ and $\overline{\nu}_e$ respectively
with
 $\nu_\mu$ and $\overline{\nu}_\mu$.
}

\label{fig:HNL_production_decay}
\end{figure}
%

In the present section we investigate the possible Majorana signatures of the type I seesaw HNLs
in experiments at $e^+ - e^-$ ($\mu^+ - \mu^-$) colliders. We consider the type I seesaw scenario with two or three
heavy Majorana neutrinos $N_{j}$, $j=1,2$ or $j=1,2,3$ discussed in some detail in the preceding section.
We assume that each $N_{j}$ is produced together with
$\bar{\nu}_{\ell'}$ or $\nu_{\ell'}$, where $\ell' = e,\mu,\tau$ or $e$
($\ell' = e,\mu,\tau$ or $\mu$),
and subsequently decays in the volume of the detector into $l^-$, and virtual or
real $W^+$, which in turn decays into $l^- + \pi^+$ if $W^+$ is virtual, or
alternatively into two identifiable jets, $W^+ \rightarrow j_1 + j_2$. Although $l^-$ can be $e^-$, $\mu^-$ or $\tau^-$, we will concentrate on the $e^-$ and $\mu^-$ options since these are easier to identify experimentally. The diagrams of the processes of interest are shown in Fig. \ref{fig:HNL_production_decay}.
The interactions generating these processes are given in
Eqs. (\ref{eq:NCC}) and (\ref{eq:NNC}). The discussion of the alternative
possibility of a Majorana $N_{j}$ decaying into $l^+ + W^-$ with
$W^- \rightarrow \tilde{j_1} +  \tilde{j_2}$ proceeds similarly to the
one we will follow below.

We will consider collider energies exceeding 10 GeV and so will neglect the masses of the initial and final state $e^\pm$ and $\mu^\pm$ as well as of the final state light neutrinos, treating $e^\pm$, $\mu^\pm$, $\bar{\nu}_{\ell'}$ and
$\nu_{\ell'}$
as massless particles. We will work also in the c.m.s. of the initial $e^+$ and $e^-$ ($\mu^+$ and $\mu^-$) and assume that each has an energy $E$, so that the initial squared invariant mass $s = 4E^2$.

We note that for sufficiently large initial energy, e.g., $s >> M^2_Z$,
the contribution of the $s$-channel $Z$-exchange diagrams to the $N_j$ production is largely subdominant and can be neglected with respect to that of the $t$-channel $W^\pm$-exchange diagrams. At the $Z$-pole, $s = M^2_Z$, however, the $Z$-decay contribution to the $N_j$ production dominates by a large margin over the $t$-channel $W$-exchange contribution.

\subsection{Single Heavy Majorana Fermion Production}
\label{sec:N}

To illustrate the main features of the processes of interest we consider first the production of a single heavy Majorana neutrino $N$, with mass $M_N$, energy $E_N$ and 3-momentum ${\bf k}_N$ at an $e^+ - e^-$ collider; the same considerations apply in the case of a $\mu^+ - \mu^-$ collider. In the c.m.s. of the colliding $e^\pm$ beams, the 3-momentum of $\nu_e$ or $\overline{\nu_e}$, ${\bf k}$, is anti-aligned with ${\bf k}_N$: ${\bf k} = -\,{\bf k}_N$. We also have:
\begin{equation}
  |{\bf k}_N| = E - \dfrac{M^2_N}{4E}\,,\quad
  E_N =  E + \dfrac{M^2_N}{4E}\,,\quad
  \beta_N =  \dfrac{E - M^2_N/(4E)}{ E + M^2_N/(4E)}\,,
  \label{eq:kNENbN}
\end{equation}
where $\beta_N$ is the $N$ velocity.

Consider the diagrams in Fig. 1 including the
$N$ and $\bar{N}$ decay parts. While the diagrams in the 1st row contribute
to the amplitude of the process of interest independently of whether $N$ is
a Dirac or Majorana fermion, the diagrams in the 2nd row, including
the $\bar{N}$ decay part, can give a nonzero
contribution only if $N$ is a Majorana particle. Thus, we refer to the former
and the latter as lepton number conserving (LNC) and lepton number violating
(LNV), respectively, although in the LNV case the lepton number violation is
``hidden'' since $\nu_l$ is not detected. Thus, in order to establish the
Majorana nature of $N$ one has to prove experimentally the presence of the
LNV term in the cross section of the process
$e^+ + e^- \rightarrow N + \overline{\nu_{\ell'}}~(\nu_{\ell'})$, $N\rightarrow l^- + W^+$,
with $W^+ \rightarrow j_1 + j_2$ (or $l^-+\pi^+$). One of the signatures of
this process is that for real $N$, $(p(l^-) + p(j_1) + p(j_2))^2
= M^2_N$ ($(p(l^-) + p(\pi^+))^2 = M^2_N$), where $p(l^-)$, $p(j_1)$ and
$p(j_2)$  ($p(l^-)$ and $p(\pi^+)$) are the 4-momenta of $l^-$, $j_1$ and
$j_2$ (of $l^-$ and $\pi^+$), respectively. For the energy of $N$ we have:
$E_N = E(l^-) + E(j_1) + E(j_2)$ ($E_N = E(l^-) + E(\pi^+)$), where the
notations are obvious.

It was shown in \cite{Petcov:1984nf} that under very general conditions
which are fulfilled in the cases of seesaw scenarios of neutrino
mass generation,  a Majorana fermion
produced via $e^+ + e^- \rightarrow N + \overline{\nu_{\ell'}}~(\nu_{\ell'})$
(or via  $\mu^+ + \mu^- \rightarrow N + \overline{\nu_{\ell'}}~(\nu_{\ell'})$)
has an angular distribution which is forward-backward (F-B) symmetric with
respect to the direction of the 3-momentum of, e.g., $e^-$, ${\bf p}_{e^-}$,
while for a Dirac fermion this symmetry does not hold.
This implies that if $N$ is produced with 3-momentum at angle
$\theta$ with respect to ${\bf p}_{e^-}$,
\begin{equation}
  \cos\theta =
  \dfrac{{\bf k}_N\cdot {\bf p}_{e^-}}{|{\bf k}_N||{\bf p}_{e^-}|}\,,
\label{eq:cth}
\end{equation}
%
the $\cos\theta$-distribution of the considered process would be symmetric with respect to $\cos\theta = 0$ if $N$ is a Majorana particle, and would not exhibit this symmetry if $N$ is a Dirac fermion. The larger part of the distribution at $\cos\theta > 0$ is due to the contribution of the LNC diagrams, while that at $\cos\theta <  0$ is generated by the LNV diagrams. Correspondingly, the presence of the F-B symmetry is due to the existence of the LNV contribution, in addition to the LNC one, in the distribution. Its absence would indicate
that $N$ is a Dirac fermion. Thus, a sufficiently precise measurement of the
$\cos\theta$- distribution of $N$ in the considered process would provide a
critical test of the nature - Majorana or Dirac - of $N$.

We note that the denominators of the t-channel $W$-boson propagators in the LNC and LNV cases, $q^2_1-M^2_W$ and $q^2_2-M^2_W$, exhibit different dependence on $\cos\theta$:
\begin{eqnarray}
    \label{eq:q1}
    q^2_1 - M^2_W &=& -\,2E\,|{\bf k}_N|\, \left (1 - \cos\theta \right) - M^2_W
    = -\,\dfrac{1}{2}(s - M^2_N)\left (1 - \cos\theta \right) - M^2_W\,,~{\rm LNC}\,,
 \\[.3cm]
 \label{eq:q2}
 q^2_2 - M^2_W &=& -\,2E\,|{\bf k}_N|\, \left (1 + \cos\theta \right) - M^2_W
= -\,\dfrac{1}{2}(s - M^2_N)\left (1 + \cos\theta \right) - M^2_W\,
 \,,~{\rm LNV}\,.
\end{eqnarray}
%
More generally, in the case of $e^+ - e^- \rightarrow N + \nu_e (\overline{\nu_e})$ ($\mu^+ - \mu^- \rightarrow N + \nu_\mu (\overline{\nu_\mu})$) we have for the $\cos\theta$-dependent factors  in the relevant differential cross sections:
\begin{eqnarray}
  \label{eq:WdiffLNC}
  \frac{d\sigma^W_{\rm LNC}}{d\cos\theta}
 & = & \dfrac{\sigma^{(0W)}_{\rm LNC}}{s}
  \dfrac{(1 + \cos\theta)\left (s + M^2_N + (s - M^2_N)\,\cos\theta\right)} 
        {\left |\dfrac{1}{2}(s - M^2_N)\,(1-\cos\theta) + M^2_W\right|^2}\,,
        \\[0.3cm]
\frac{d\sigma^W_{\rm LNV}}{d\cos\theta}
         & = & \dfrac{\sigma^{(0W)}_{\rm LNV}}{s}
         \dfrac{(1 - \cos\theta)
    \left (s + M^2_N - (s - M^2_N)\,\cos\theta\right)} 
         {\left |\dfrac{1}{2}(s - M^2_N)\,(1 + \cos\theta) + M^2_W\right|^2}\,.
\label{eq:WdiffLNV}
\end{eqnarray}
%
Here $d\sigma^W_{\rm LNC(LNV)}/d\cos\theta$ is the contribution to the
differential cross section of the LNC (LNV) $t$-channel $W^\pm$-exchange diagram
and $\sigma^{(0W)}_{\rm LNC(LNV)}$ is the factor in the cross section that does
not depend on $\cos\theta$ that will be given below
\footnote{The same result is valid if taking into account the decay of
real $N\rightarrow l^- + W^+$ with real or virtual $W^+$. However, 
$\sigma^{(0W)}_{\rm LNC(LNV)}$ depends on whether $W^+$ is real or virtual.}.
Thus, $d\sigma^W_{\rm LNC(LNV)}/d\cos\theta$ has an absolute maximum at
$\theta = 0~(\pi)$, with $d\sigma^W_{\rm LNV(LNC)}/d\cos\theta = 0$ at
$\theta = 0~(\pi)$. The discussed LNC (LNV) contribution to the $\cos\theta$ distribution is dominant in the $\cos\theta > 0$ ($\cos\theta < 0$) region of the distribution \footnote{In the case of $N\rightarrow l^+ + W^-$ decay,
$d\sigma^W_{\rm LNC(LNV)}/d\cos\theta$ has an absolute maximum at
$\theta = \pi~(0)$ and is dominant in the region $\cos\theta < 0$
($\cos\theta > 0$).}. At $\theta = \pi/2$ we have
$d\sigma^W_{\rm LNC}/d\cos\theta = d\sigma^W_{\rm LNV}/d\cos\theta$. 

Similar considerations apply to the LNV and LNC $s$-channel $Z$-boson diagram contributions to the $\cos\theta$-distribution of the considered process with real $N$ decaying into $l^-$ and virtual/real $W^+$, which in turn gives two jets or $l^- + \pi^+$ in the final state. Indeed, as can be shown,
\begin{eqnarray}
  \label{eq:ZLNC}
  \frac{d\sigma^Z_{\rm LNC}}{d\cos\theta}
  & = & \dfrac{\sigma^{(0Z)}_{\rm LNC}}{\cos^4\theta_W}\,
  \dfrac{ g^2_L (1+\cos\theta)^2 + g^2_R (1-\cos\theta)^2 +
    (g^2_L +  g^2_R)\dfrac{M^2_N}{s}\sin^2\theta}
{\left |s - M^2_Z + i\,\Gamma_Z\,M_Z\right|^2}\,,\\[0.3cm]
\label{eq:ZLNV}
  \frac{d\sigma^Z_{\rm LNV}}{d\cos\theta}
  & = & \dfrac{\sigma^{(0Z)}_{\rm LNV}}{\cos^4\theta_W}
  \dfrac{g^2_L (1 - \cos\theta)^2 + g^2_R (1 + \cos\theta)^2 +
    (g^2_L +  g^2_R)\dfrac{M^2_N}{s}\sin^2\theta}
{\left |s - M^2_Z + i\,\Gamma_Z\,M_Z\right|^2}\,,
\end{eqnarray}
%
$d\sigma^Z_{\rm LNC(LNV)}/d\cos\theta$ being the contribution to the differential
cross section of the LNC (LNV) $s$-channel $Z$-exchange diagram,
$\sigma^{(0Z)}_{\rm LNC(LNV)}$ is a $\cos\theta$-independent factor in the cross
section, $g_L = -\,1/2 + \sin^2\theta_W$ and $g_R = \sin^2\theta_W$ are the SM chiral
$e^\pm$ ($\mu^\pm$) couplings to $Z$, with $\sin^2\theta_W = 0.22348$~\cite{ParticleDataGroup:2024cfk}.

It is possible to derive an analytic expression for
$\sigma^{(0W,0Z)}_{\rm LNC(LNV)}$.
Anticipating the discussion of the cases of seesaw scenarios with two and
three RH neutrinos, we give below the expression for a virtual $N$ and,
for simplicity, for real $W^+$ and $l^-$ produced in the decay
$N(\bar{N})\rightarrow l^-+ W^+$:
\begin{eqnarray}
  \nonumber
  \sigma^{(0W,0Z)}_{\rm LNC(LNV)} &=&
  \left( \dfrac{G_F}{\sqrt{2}}\right )^3
  \dfrac{M^4_W}{16\pi^2\,s}\,\int_{M_W^2}^{s}
  \dfrac{dq_N^2}{q_N^2}\,\sigma^{(0)}(q_N^2)\,|F^{W,Z}_{\rm LNC(LNV),j}(q_N^2)|^2\,,
  \\[0.25cm]
  \\
\sigma^{(0)}(q_N^2) &=& \left(s-q_N^2\right)^2\,\left(q_N^2-M_W^2\right)^2\,
\left(q_N^2+2M_W^2\right)\,,~~
\label{eq:sigmaWZ0}
\end{eqnarray}
%
where $q_N^2$ is the 4-momentum squared of $N$ and 
\begin{eqnarray}
  \label{eq:FW}
&F^{W}_{\rm LNC,j}(q_N^2) =\dfrac{ (RV)_{\ell j}(RV)^*_{e j}}
           {q_N^2-M_j^2+iM_j\Gamma_j}\,,~  
F^{W}_{\rm LNV,j}(q_N^2) = \dfrac{\rho_j\,M_j(RV)_{\ell j}(RV)_{e j}}
{\sqrt{q_N^2}(q_N^2-M_j^2+iM_j\Gamma_j)}\,,~j\equiv {\rm N}\,,\\[0.30cm]
&F^{Z}_{\rm LNC,j}(q_N^2) =\dfrac{(RV)_{\ell j}
             \sum_{\ell'=e,\mu,\tau}(RV)^*_{\ell' j}}
           {q_N^2-M_j^2+iM_j\Gamma_j}\,,~
F^{Z}_{\rm LNV,j}(q_N^2) = \dfrac{\rho_j\,M_j(RV)_{\ell j}
             \sum_{\ell'=e,\mu,\tau}(RV)_{\ell' j}}
           {\sqrt{q_N^2}(q_N^2-M_j^2+iM_j\Gamma_j)}\,,~~
           j\equiv {\rm N}.
           \label{eq:FZ}
\end{eqnarray}
%
Here  $\Gamma_N$ is the total width of $N$
\footnote{We note that in the case of identical charged and neutral current
couplings considered by us, $\Gamma^M_N = 2\Gamma^D_N$, $\Gamma^{M(D)}_N$ being
the total width of a Majorana (Dirac) HNL $N$.},
$(RV)_{\ell N}$ ($(RV)_{\ell' N}$) is  the coupling of $N$ to the
charged lepton $\ell$ ($\ell'$)  (flavour neutrino $\nu_\ell$ ($\nu_{\ell'}$)) 
in the weak charged (neutral) lepton current
(see Eqs. (\ref{eq:NCC}) and (\ref{eq:NNC})),
$\rho_N = \pm 1$ being a sign factor in the Majorana condition for $N$, $C (\overline{N})^T = \rho_N\, N$ (see, e.g., \cite{Bilenky:1987ty}). In the case of
a single Majorana $N$, the presence of $\rho_N$ in Eqs. (\ref{eq:FW})
and (\ref{eq:FZ})
has no physical implications. The case of real $N$ can be obtained by setting
$q_N^2 = M_N^2$ and replacing $|q_N^2-M_N^2+iM_N\Gamma_N|^{-2}$ with
$\pi\,\delta(q_N^2-M_N^2)/(M_N\Gamma_N)$ (narrow width approximation).
We note that the same factor $|F^{W(Z)}_{\rm LNC(LNV),j}(q_N^2)|^2$
is present in $d\sigma^{W(Z)}_{\rm LNC(LNV)}/d\cos\theta$
also when the $W^+$ produced in the decay of $N$ ($\bar{N}$) is virtual.

In the general case when both the $W^\pm$- and $Z$- exchange contributions
are taken into account, the differential cross section of production of a
Majorana fermion $N$ via the process
$e^+ + e^- \rightarrow N + \overline{\nu_{\ell'}}~(\nu_{\ell'})$,
with $N \rightarrow l^- + W^+$ with virtual or real $W^+$, then is given by:
\begin{equation} 
  \frac{d\sigma_{\rm M}}{d\cos\theta}
  =  \frac{d\sigma_{\rm LNC}(|A_W + A_Z|^2)}{d\cos\theta} +
  \frac{d\sigma_{\rm LNV}(|A_W + A_Z|^2)}{d\cos\theta}\,,
  \label{eq:MajNprod}
\end{equation}
%
while for a Dirac $N$ the cross section reads:
\begin{equation} 
  \frac{d\sigma_{\rm D}}{d\cos\theta}
  =  \frac{d\sigma_{\rm LNC}(|A_W + A_Z|^2)}{d\cos\theta}\,.
  \label{eq:DNprod}
\end{equation}
%
Here $d\sigma_{\rm LNC(LNV)}(|A_W + A_Z|^2)/d\cos\theta$ is the differential cross section including the $W$- and $Z$- exchange contributions. As is indicated by the notation, the interference of the $W$- and $Z$- exchange diagram
contributions in the amplitude of the considered process
has to be taken into account in the calculation of
$d\sigma_{\rm LNC(LNV)}(|A_W + A_Z|^2)/d\cos\theta$ when neither of the two
contributions is strongly dominating.

One can use instead of $\cos\theta$ the rapidity of $N$, $y_N$, to describe
the process of interest \cite{arXiv:2403.06561}
(see also \cite{Hernandez:2018cgc}):
\begin{equation}
  y_N = \dfrac{1}{2}\, \ln \dfrac{E_N + |{\bf k}_N|\cos\theta}
  {E_N - |{\bf k}_N|\cos\theta}
= \dfrac{1}{2}\, \ln \dfrac{1 + \beta_N \,\cos\theta}
  {1 - \beta_N\,\cos\theta}\,.
  \label{eq:yN}
\end{equation}
%
We note that $y_N > 0$ ($y_N < 0$)  for $\cos\theta >  0$ ($\cos\theta <  0$), and that $y_N = 0$ at  $\cos\theta =  0$. Therefore the forward-backward symmetry in the $\cos\theta$-distribution corresponds to a symmetry between the distributions at $y_N > 0 $ and at  $y_N < 0 $. The $y_N$-distribution would be symmetric with respect to $y_N = 0$ for a Majorana $N$ and would not have this symmetry if $N$ is a Dirac particle. As can be shown, all features of the $\cos\theta$-distribution discussed above are present also in the $y_N$-distribution. In particular, in the case of  $N \rightarrow l^- + W^+$ decay considered by us, the LNC (LNV) contribution to the $y_N$ distribution has an absolute maximum and is dominant in the $y_N >  0$ ($y_N < 0$) region of the distribution.

\begin{figure}
    \centering
    \includegraphics[width=0.99\linewidth]{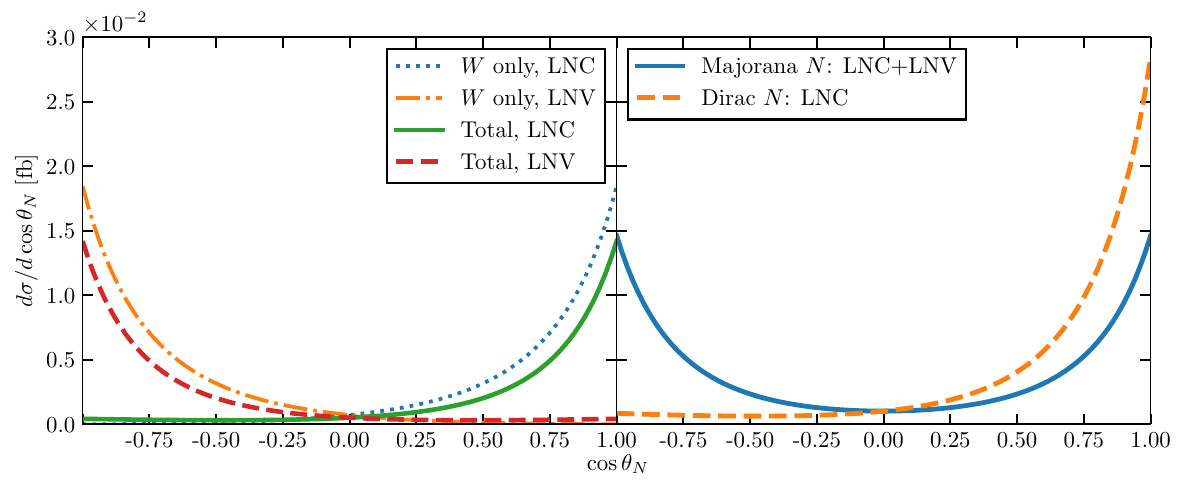}
    \caption{The differential cross sections
      $d\sigma^W_{\rm LNC,LNV}/d\cos\theta$ and
$d\sigma_{\rm LNC,LNV}(|A_W + A_Z|^2)/d\cos\theta$
(left panel), and 
$d\sigma_{\rm M,D}/d\cos\theta$ (right panel)
 for $M_N = 60$ GeV and $\sqrt{s} = 160$ GeV. See text for further details.}
    \label{fig:M60s160}
\end{figure}
\begin{figure}
    \centering
 \includegraphics[width=0.96\linewidth]{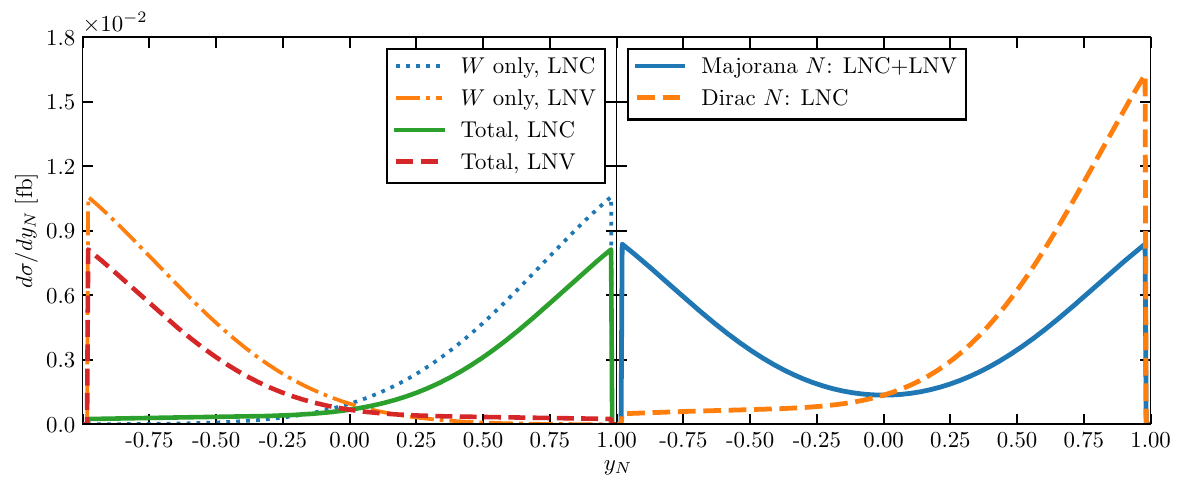}
    \caption{The differential cross sections
      $d\sigma^W_{\rm LNC,LNV}/dy_N$ and
$d\sigma_{\rm LNC,LNV}(|A_W + A_Z|^2)/dy_N$
(left panel), and 
$d\sigma_{\rm M,D}/dy_N$ (right panel)
 for $M_N = 60$ GeV and $\sqrt{s} = 160$ GeV. See text for further details.}
    \label{fig:M60s160y}
\end{figure}

We show in Fig. \ref{fig:M60s160} the cross sections 
$d\sigma^W_{\rm LNC,LNV}/d\cos\theta$ and
$d\sigma_{\rm LNC,LNV}(|A_W + A_Z|^2)/d\cos\theta$
(left panel) and 
$d\sigma_{\rm M,D}/d\cos\theta$ (right panel)
for $M_N = 60$ GeV and $\sqrt{s} = 160$ GeV.
The decay $N\rightarrow \mu^- + W^+$ with virtual $W^+$ is assumed.
The values of $N$ couplings used are $|(RV)_{eN}| = |(RV)_{\mu N}| = 10^{-3}$
and $(RV)_{\tau N} = 0$.
The final state $W$-boson is considered to be off-shell
producing two distinct jets in the final state.
 As indicated in the left panel, there is a non-negligible contribution
 of the $W$- and $Z$-exchange interference term in 
$d\sigma_{\rm LNC,LNV}(|A_W + A_Z|^2)/d\cos\theta$.
 All the features of the LNC and LNV contributions discussed earlier in
 this subsection are present in the left panel of
 the figure. The maximum at $\cos\theta > 0$ corresponds to the
 contribution of the  LNC diagram, while that symmetric with respect to
 $\cos\theta =0 $ at $\cos\theta < 0$ is due to the LNV diagram.
 As seen in the right panel,
 the production of the Majorana
 $N$ is perfectly $F-B$ symmetric, while that of the Dirac $N$  is strongly
 $F-B$ asymmetric, with an integrated $F-B$ asymmetry $A_{FB} = 0.811$
 \footnote{$A_{FB}$ is defined as
 $\int_{0}^{1}d\sigma_{\rm D}/d\cos\theta - \int_{-1}^{0}d\sigma_{\rm D}/d\cos\theta$
 divided by the sum of the two integrals.}.
 For fixed $M_N = 60$ GeV the asymmetry increases with $\sqrt{s}$ 
 and at $\sqrt{s} = 240$ GeV (350 GeV)  we have $A_{FB} = 0.952~(0.982)$.

 As shown in Fig. \ref{fig:M60s160y} obtained for the same values
  of  $M_N$, $\sqrt{s}$ and $|(RV)_{lN}|$,
analogous behavior is exhibited by the rapidity distribution
 of $d\sigma_{\rm LNC,LNV}(|A_W + A_Z|^2)/dy_N$
(left panel) and 
$d\sigma_{\rm M,D}/dy_N$ (right panel).
 The presence of the LNV contribution leads to a distinct
 maximum at $y < 0$ in addition to the one at $y > 0$ due to the
 LNC contribution, and to
 $A_{FB} = 0$ if $N$ is a Majorana fermion. 
We note also that for real $N$, the considered cross sections
are proportional to the $N\rightarrow \mu^- + j_1 + j_2$ decay
branching ratio and therefore scale with the $N$ couplings practically 
only as  $|(RV)_{eN}|^2$ (and not as $|(RV)_{eN}(RV)_{\mu N}|^2$).

\begin{figure}
    \centering
    \includegraphics[width=0.99\linewidth]{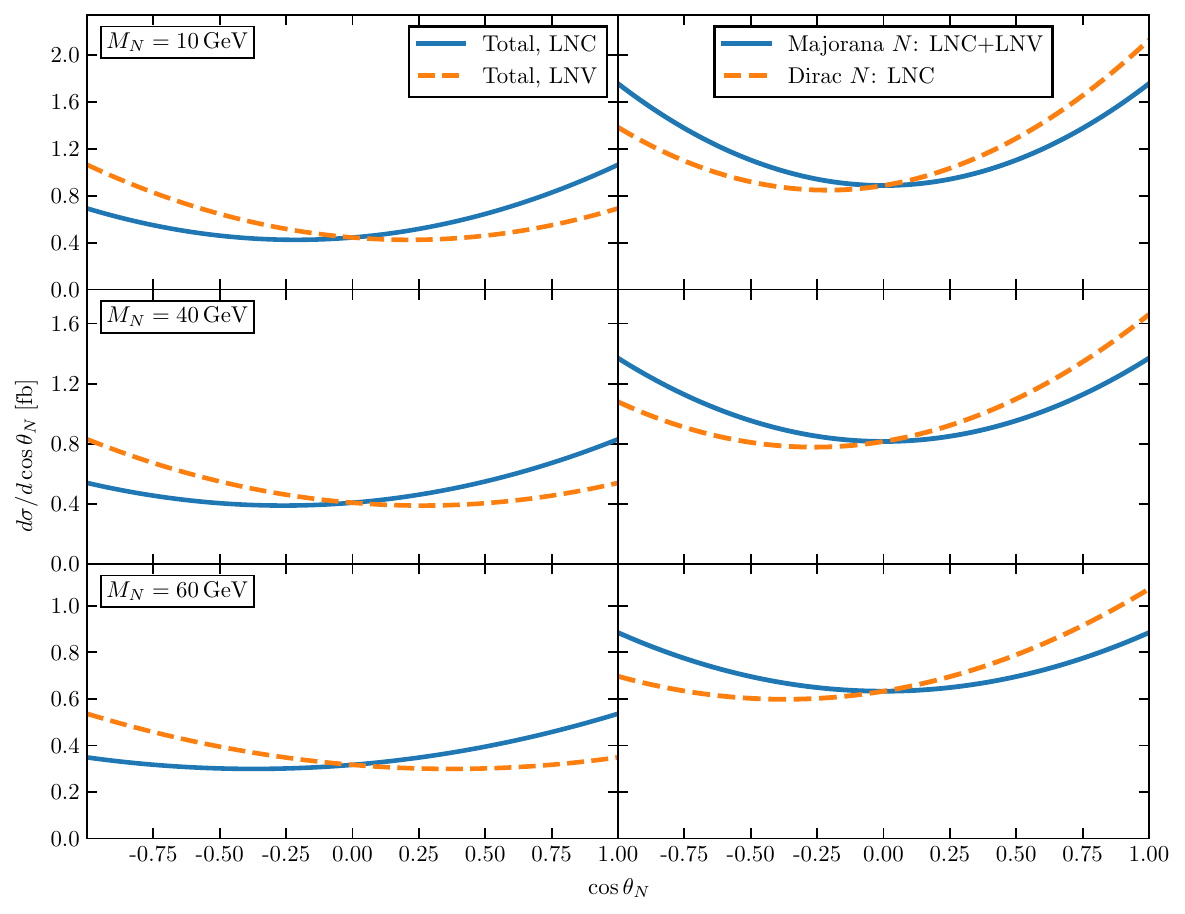} 
    \caption{$d\sigma^Z_{\rm LNC,LNV}/d\cos\theta$ and
    $d\sigma_{\rm LNC,LNV}(|A_W + A_Z|^2)/d\cos\theta$
    (left panels), and 
    $d\sigma_{\rm M,D}/d\cos\theta$ (right panels) for $N$ with mass 
      $M_N = 10$, 40 and 60 GeV produced at the $Z$-pole, 
      $\sqrt{s} = M_Z = 91.1876$ GeV. 
      See text for further details.}
    \label{fig:NatZpole}
\end{figure}
%

Similar behavior is seen in Fig. \ref{fig:NatZpole} showing
the cross sections 
$d\sigma^Z_{\rm LNC,LNV}/d\cos\theta$ and
$d\sigma_{\rm LNC,LNV}(|A_W + A_Z|^2)/d\cos\theta$
(left panels) and 
$d\sigma_{\rm M,D}/d\cos\theta$ (right panels)
for production of $N$ with mass $M_N = 10$, 40 and 60 GeV 
at the $Z$-pole (corresponding to the first phase of FCCee).
As in  Fig. \ref{fig:M60s160}, 
the $N\rightarrow \mu^- + W^+ \rightarrow \mu^- + j_1 + j_2$ decay is assumed
and the same values of the relevant $N$ couplings are used: 
$|(RV)_{eN}| = |(RV)_{\mu N}| = 10^{-3}$, $(RV)_{\tau N} = 0$.
The $F-B$ asymmetry in the case of Dirac $N$ is noticeably smaller
than that shown in Fig. \ref{fig:M60s160}, right panel,
and decreases somewhat with $M_N$: we find $A_{FB} = 0.158$, 0.145, and 0.131
for $M_N = 10$, 40 and 60 GeV, respectively.

%
\subsection{Production of the Seesaw Heavy Majorana Neutrinos}
\label{sec:Ni}

We consider the seesaw scenarios with two or three RH neutrinos discussed
in Section \ref{sec:spseesaw}.
We focus our discussion on the LNV contribution
to the amplitude of the process of interest, which we consider for
concreteness to be 
$e^+ + e^- \rightarrow N_j + \bar{\nu}_{\ell'}~(\bar{N}_j + \nu_{\ell'})$,
$N_j~(\bar{N}_j) \rightarrow l^- + W^+$,
$W^+ \rightarrow j_1 + j_2$, $j = 1,2$ or $j=1,2,3$,
$\ell'=e,\mu,\tau$ or $e$,
$\ell^- = e^-$ or $\mu^-$,
where $N_j \equiv \bar{N}_j$ is a real or virtual heavy Majorana neutrino,
$W^+$ can be real or virtual and the decay of 
$N_j$ is assumed to take place in the volume
of the detector. The factors in the LNC and LNV contributions to the amplitude
of this process, originating from $W$- and $Z$- exchange diagrams,
which are of interest for our analysis read:
\begin{equation}
  P^{W}_{\mathrm{LNC(LNV)},k} = \sum_{j=1}^{k}  F^{W}_{{\rm LNC(LNV)},j}(q_N^2)\,,~~
P^{Z}_{\mathrm{LNC(LNV)},k} =\sum_{j=1}^{k}  F^{Z}_{{\rm LNC(LNV)},j}(q_N^2)\,,~k=2~{\rm or}~3\,.
\label{eq:P23WZ}
\end{equation}
%
Here $q_N$ is the 4-momentum of the 
real or virtual $N_j$,
$M_j$, $\Gamma_{j}$ and $\rho_j$ are the mass, 
the width of $N_{j}$ and the sign factor in the Majorana condition
satisfied by $N_{j}$, and $F^{W,Z}_{\rm LNC(LNV),j}(q_N^2)$ are given in
Eqs. (\ref{eq:FW}) and (\ref{eq:FZ}). 

%
\subsubsection{The Seesaw Scenarios with Two RH Neutrinos}
%

These are scenarios with two heavy Majorana neutrinos $N_{1,2}$ with masses
$M_{1,2}$,  $M_1 < M_2$. If in the process of interest, e.g.,  $N_2$ is
on mass shell, the invariant mass of the final state charged
lepton $l^-$ and the two jets should be equal to the mass of $N_2$.
It is important to note that $N_{1,2}$ are not directly observed.
The observation of the process of interest 
with the characteristic Breit-Wigner enhancement of the cross section
due to the $N_2$ propagator when the invariant
mass of the two jets and $l^-$ approaches the mass of $N_2$,
would be a signature of the Majorana nature
of $N_2$ if in addition it is established that $A_{FB} = 0$.
However, the factors of the LNV contribution
to the amplitude of the discussed process which are 
relevant for the determination of the Majorana nature of
$N_{2}$ have the form:
\begin{eqnarray}
\nonumber
\sqrt{q^2_N} P^Z_{\mathrm{LNV},2} &=& \dfrac{M_2 (RV)_{\ell 2}\sum_{\ell'}(RV)_{\ell' 2}}
{q_N^2-M_2^2+ iM_2\Gamma_2} -\,
 \dfrac{M_1 (RV)_{\ell 1}\sum_{\ell'}(RV)_{\ell' 1}}
           {q_N^2-M_1^2+iM_1\Gamma_1}\,,\\[0.30cm] 
\sqrt{q^2_N}P^W_{\mathrm{LNV},2} &=& \dfrac{M_2 (RV)_{\ell 2}(RV)_{e 2}}
{q_N^2-M_2^2+iM_2\Gamma_2} -\,
 \dfrac{M_1 (RV)_{\ell 1}(RV)_{e 1}}
           {q_N^2-M_1^2+iM_1\Gamma_1}\,, 
\label{eq:PWZ20}           
\end{eqnarray}
%
where we took into account the fact that in the scenarios we
consider $\rho_1 = -\,1$ and $\rho_2 = 1$.
In the cases of interest $M_{1,2} \gtap 10$ GeV and $M_{1,2} < M_W$
($M_{1,2} > M_W$),
$\Gamma_{1 (2)}\propto G_F^2 M^5_{1 (2)} \ll M_{1 (2)}$
($\Gamma_{1 (2)}\propto G_F M^3_{1 (2)} \ll M_{1 (2)}$).
We note that the above expressions are valid for any value of 
the invariant mass $q^2_N$. It should be emphasised that 
even when $N_{2}$ is on mass shell, $q^2_N = M^2_{2}$,
and the first term in Eq. (\ref{eq:PWZ20}) dominates 
due to the fact that, e.g.,  $M_2$ is much 
bigger than $M_1$, the contribution of the virtual 
$N_1$ (i.e., the second term in Eq. (\ref{eq:PWZ20})) 
is always present in the amplitude. 

In the symmetry-protected scenario with 
exact conservation of the non-standard lepton charge $L^\prime$,
as we have shown (see Eqs. (\ref{eq:Lsym})- (\ref{eq:Vsym}) and the
related discussion),
the fields $N_{1,2}$ form a Dirac field $N_D$, they couple to the
$W^\pm$ and $Z^0$ bosons only through the LH component of $N_D$ and
there are no physical observable effects of the Majorana nature of $N_{1,2}$.
Indeed, we have $(RV)_{l2} = -\,(RV)_{l1}$, $M_1 = M_2$, $\Gamma_1 = \Gamma_2$,
$P^{Z,W}_{LNV,2} = 0$ and, as a consequence, the LNV contribution vanishes.

\vspace{0.5cm}
{\bf Scenario with $L^\prime$-symmetry breaking by neutrino Yukawa coupling}

\vspace{0.3cm}
Consider next the scenario with $L^\prime$ breaking originating
from the neutrino Yukawa coupling analysed in 
subsection \ref{sec:LpB1}. In this case $M_1\neq M_2$, but 
the difference between the masses of
$N_1$ and $N_2$ is exceedingly small: for NH light neutrino mass spectrum
$M_2 - M_1 = m_3 - m_2 \cong 0.0414$ eV, where $m_{2,3}$ are the two
non-zero light neutrino masses. 
Now $N_{1,2}$ form a pseudo-Dirac pair 
$N_{PD} = (N_{1} + N_{2})/\sqrt{2}$.
Since $M_2 - M_1$ is exceedingly small, the observation
of each of the two $N_{1,2}$ via the
corresponding Breit-Wigner resonance enhancement of
$q^2_N$-dependence of the cross section of the process 
$e^+ + e^- \rightarrow l^- + j_1 + j_2 + E_{mis}$,
where $q^2_N = (p(l^-) + p(j_1) + p(j_2))^2 \equiv M^2_{l^-j_1j_2}$ and 
$E_{mis}$ stands for ``missing energy'', $l^-=e^-,\mu^-$, at
$\sqrt{q^2_N} = M_{l^-j_1j_2} = M_1$ and at $M_{l^-j_1j_2} = M_2$,
seems hardly possible. 

Requiring that the $N_{1,2}$ CC and NC couplings are sizeable
leads in the example considered in subsection \ref{sec:LpB1}
to $(RV)_{l1} = -\, (RV)_{l2} (1 + O(10^{-7}))$,
$|(RV)_{l1,2}| \lesssim 7\times 10^{-3}\,(10~{\rm GeV}/M_{12})$.
Thus, $(RV)_{l1,2}$ can be relatively large implying the possibility
of $N_{1,2}$ production at $e^+ - e^-$ (or  $\mu^+ - \mu^-$) colliders
with observable rates.

In the considered scenario the $P^{Z,W}_{\mathrm{LNV},2}$ factors are given by:
\begin{eqnarray}
    \nonumber
\sqrt{q^2_N}P^Z_{\mathrm{LNV},2} &=&\left [
  \dfrac{M_2 }
{q_N^2-M_2^2+iM_2\Gamma_2} -\, \dfrac{M_1}{q_N^2-M_1^2+iM_1\Gamma_1}\right ]\,
(RV)_{\ell 2}\sum_{\ell'}(RV)_{\ell' 2}\,,\\[0.25cm]
\sqrt{q^2_N} P^W_{\mathrm{LNV},2} &=&\left [
  \dfrac{M_2 }
{q_N^2-M_2^2+iM_2\Gamma_2} -\,
 \dfrac{M_1}
           {q_N^2-M_1^2+iM_1\Gamma_1}\right ](RV)_{\ell 2}(RV)_{e 2}\,. 
\label{eq:P2WZLBY}           
\end{eqnarray}
%
Thus, $P^{W,Z}_{\mathrm{LNV},2} \propto (M_2 - M_1)$ and the LNV contributions to the
cross section of the process of interest seem to be strongly suppressed.
At the same time the LNC contributions can be relatively
large and the production of $N_{1,2}$ can take place with observable rates.
The $P^{Z}_{\mathrm{LNC},2}$ factor, for example,  has the form:
\begin{equation}
P^Z_{\mathrm{LNC},2} = \left [
  \dfrac{1}
{q_N^2-M_2^2+iM_2\Gamma_2} +\, \dfrac{1}{q_N^2-M_1^2+iM_1\Gamma_1}\right ]\,
(RV)_{\ell 2}\sum_{\ell'}(RV)_{\ell' 2}\,.
\label{eq:PZLNC2MD}
\end{equation}
%
Thus, for the ratio of $P^{Z}_{\mathrm{LNV},2}$ and $P^Z_{\mathrm{LNC},2}$ we get:
\begin{equation}
  \dfrac{P^{Z}_{\mathrm{LNV},2}}{P^Z_{\mathrm{LNC},2}} =
  \dfrac{1}{\sqrt{q^2_N}}
  \dfrac{(M_2-M_1)\left[q^2_N + M_1M_2 -
      i\dfrac{M_1M_2}{M_2-M_1}(\Gamma_2 - \Gamma_1)\right]}
        {2q^2_N - (M^2_1 + M^2_2) + i(M_1\Gamma_1 + M_2\Gamma_2)}\,.
        \label{eq:PZLNV2dPZLNC2}
\end{equation}
%
The Majorana nature of $N_{1,2}$ might be observable only if
$|P^{Z}_{\mathrm{LNV},2}/P^Z_{\mathrm{LNC},2}|^2 \sim 1$; for 
$|P^{Z}_{\mathrm{LNV},2}/P^Z_{\mathrm{LNC},2}|^2 \ll 1$, the production of $N_{1,2}$ would mimic
that of a Dirac fermion (see also \cite{Hernandez:2018cgc}).
We consider next three examples which would be sufficient
to illustrate the difficulties of probing the Majorana nature of $N_{1,2}$
and relating them to a seesaw scenario.

Assume first that $M_1 \cong M_2 = 60$ GeV with
$M_2 - M_1 \cong 0.0414$ eV and
$((RV)_{l1,2})^2 \cong 2.5\times 10^{-7}$, $l=e,\mu,\tau$.
The widths of $N_{1,2}$ in the considered case
are given by:
\begin{equation}
\Gamma_j = 5.92\times 10^{-6}\,M_j \left (\dfrac{M_j}{60~{\rm GeV}}\right)^{4}\,
\sum_l|(RV)_{lj}|^2\,,~~j=1,2\,.
\label{eq:Gj}
\end{equation}
%
With the chosen values of $M_{1,2}$ and $(RV)_{l1,2}$ we find
for $q^2_N \equiv M^2_{l^-j_1j_2}=  M^2_1,M^2_2$:
$|\Gamma_{2} -\Gamma_{1}|\ll \Gamma_{1,2}$ and 
$\Gamma_{1,2} \cong 0.266~{\rm eV} \gg (M_2-M_1)$.
Correspondingly,
$|P^{Z}_{\mathrm{LNV},2}/P^Z_{\mathrm{LNC},2}|^2 \cong (M_2-M_1)^2/\Gamma^2_{1,2} \ll 1$.
This implies that the production of $N_{1,2}$ at the $Z$-pole will mimic that
of a single Dirac fermion. In this case one would have $A_{FB}\neq 0$.
\begin{figure}
    \centering
    \includegraphics[width=0.99\linewidth]{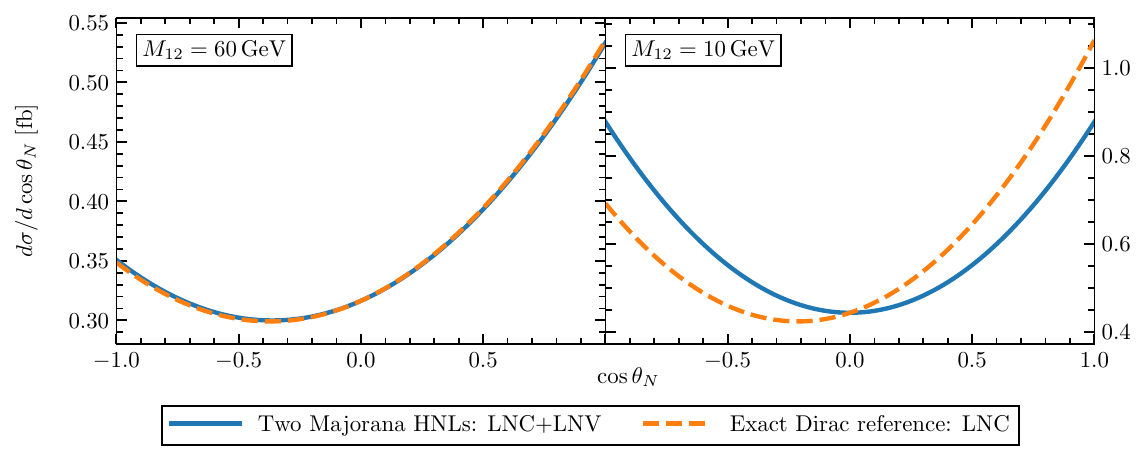}
    \caption{The differential cross sections
        $d\sigma_{\rm M,D}/d\cos\theta$
        for the production of $N_{1,2}$ at the $Z$-pole,
        $\sqrt{s}=M_Z=91.1876$ GeV, with
        $((RV)_{l1,2})^2 \cong 2.5\times10^{-7}$, $l=e,\mu,\tau$.
        In the left panel,
        $M_1 \cong M_2 = 60$ GeV, with
        $M_2-M_1 \cong 0.0414$ eV.
        In the right panel,
        $M_1 \cong M_2 = 10$ GeV, with
        $M_2-M_1 \cong 0.0414$ eV.
        See text for further details.
    }
    \label{fig:YukawaBreakingZpole}
\end{figure}
%
This is illustrated in the left panel of Fig.~\ref{fig:YukawaBreakingZpole} which shows the
$\cos\theta$-distribution of $N_{1,2}$ generation at the $Z$-pole for
$M_{12} = 60$ GeV. Since $A_{FB} \cong 0.127 \neq 0$, which is 
very close to the exact Dirac value of $A_{FB} \cong 0.130$,
it clearly mimics the production of a single Dirac fermion $N_D$
with a mass of $M_{12} = 60$ GeV. The mean free paths of $N_{1,2}$
($N_D$) are approximately $d_{1,2}\cong 3.2\times 10^{-5}$ cm,
so $N_{1,2}$ will decay within the detector.

Consider next the case of 
$M_1 \cong M_2 = 10$ GeV with the same values of
$|(RV)_{l1,2}|^2 \cong 2.5\times10^{-7}$, $l=e,\mu,\tau$.
Using Eq. (\ref{eq:Gj}) we find
$\Gamma_{1,2} \cong 3.43\times 10^{-5}~{\rm eV} \ll (M_2 - M_1)$,
$|\Gamma_{2} - \Gamma_{1}| \ll \Gamma_{1,2}$.
For the ratio of interest we obtain
at $q^2_N \equiv M^2_{l^-j_1j_2} = M^2_1,M^2_2$: $|P^{Z}_{\mathrm{LNV},2}/P^Z_{\mathrm{LNC},2}| \cong 1$.
As a consequence, the $\cos\theta$-dependence of the production of
$N_{1,2}$ at the $Z$-pole will exhibit a forward-backward symmetry -
the indicator of the Majorana particle production,
as the results of a numerical analysis shown in the right panel of Fig.~\ref{fig:YukawaBreakingZpole}
confirm. It is quite remarkable that the F-B symmetry indicator allows us to reveal
the Majorana nature of $N_{1,2}$ even if in the considered case
$(M_2-M_1)/M_{1,2} \cong 4\times 10^{-12}$. 
However,
since  $(M_2 - M_1)$ is exceedingly small, the observation
of each of the two $N_{1,2}$ via the
corresponding Breit-Wigner resonance enhancement of the
$q^2_N$-dependence of the cross section of the process 
$e^+ + e^- \rightarrow l^- + j_1 + j_2 + E_{mis}$,
where $q^2_N = (p(l^-) + p(j_1) + p(j_2))^2 \equiv M^2_{l^-j_1j_2}$ and 
$E_{mis}$ stands for ``missing energy'', $l^-=e^-,\mu^-$, at
$\sqrt{q^2_N} = M_{l^-j_1j_2} = M_1$ and at $M_{l^-j_1j_2} = M_2$,
seems hardly possible. If it is indeed impossible
to disentangle the presence of
$N_1$ from that of $N_2$ in the process of interest,
the production of $N_{1,2}$
would mimic that of a single Majorana fermion.
\begin{figure}[t]
    \centering
 \includegraphics[width=0.96\linewidth]{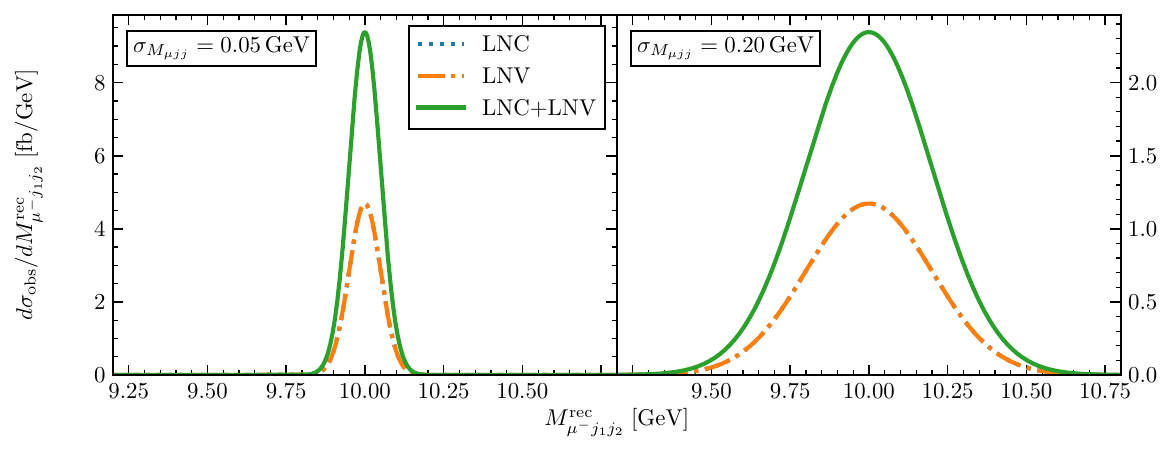}
 \caption{The differential cross sections
     $d\sigma(e^+ + e^- \rightarrow \mu^- + j_1 + j_2 + E_{mis})/dM_{\mu^-j_1j_2}$
      at the $Z$-pole,
      $\sqrt{s} = M_Z = 91.1876$ GeV, for 
      $M_1 \cong M_2 = 10$ GeV, $M_2 - M_1 \cong 0.0414$ eV,
        $((RV)_{l1,2})^2 \cong 2.5\times 10^{-7}$, $l=e,\mu,\tau$, assuming
      the measured $M_{\mu^-j_1j_2}$ follows the Gaussian distribution with
      a $1\sigma$ error of $\delta M_{l^-j_1j_2} =  0.05$ GeV and 0.20 GeV.
    See text for further details. 
 }
    \label{fig:2RHZpoleN12Gauss}
\end{figure}
%
This possibility is illustrated graphically in
Fig. \ref{fig:2RHZpoleN12Gauss} which shows 
the differential cross sections
$d\sigma(e^+ + e^- \rightarrow \mu^- + j_1 + j_2 + E_{mis})/dM_{\mu^-j_1j_2}$
at the $Z$-pole, $\sqrt{s} = M_Z$, as a function of $M_{\mu^-j_1j_2}$ for 
      $M_1 \cong M_2 = 10$ GeV, $M_2 - M_1 \cong 0.0414$ eV,
$((RV)_{l1,2})^2 \cong 2.5\times 10^{-7}$, $l=e,\mu,\tau$, assuming
the measured $M_{\mu^-j_1j_2}$ follows the Gaussian distribution with
 a $1\sigma$ error of $\delta M_{l^-j_1j_2} =  0.05$ GeV and 0.20 GeV.
Since $\delta M_{l^-j_1j_2} \gg M_2 - M_1$, 
the Breit-Wigner resonance enhancement of the cross section,
clearly seen in the figure, corresponds to a production of a single fermion
with a mass of 10 GeV. We note that the mean free paths of $N_{1,2}$
are approximately $d_{1,2}\cong 2.59$ cm,
so $N_{1,2}$ will decay within the detector with
the displaced decay vertex signature.

\begin{figure}
    \centering
    \includegraphics[width=0.70\linewidth]{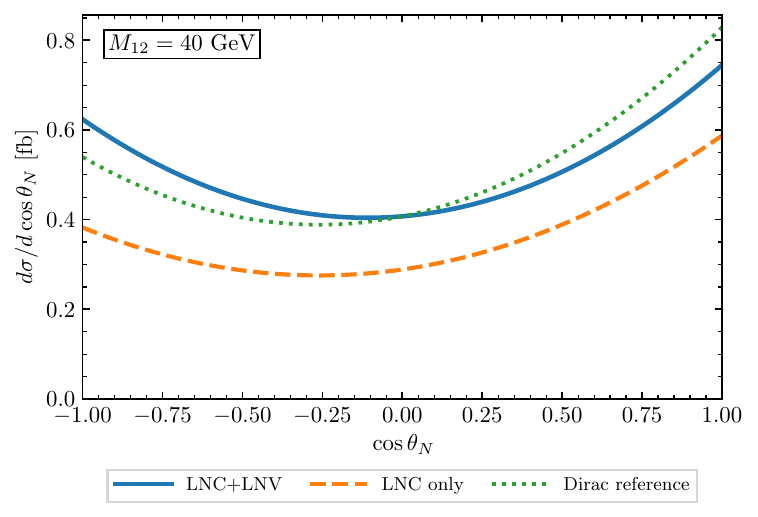}
    \caption{The differential cross sections
        $d\sigma_{\rm M,D}/d\cos\theta$
      for the production of $N_{1,2}$ with
      $M_1 \cong M_2 = 40$ GeV, 
      $M_2-M_1 \cong 0.0414$ eV, and
      $((RV)_{l1,2})^2 \cong 2.5\times10^{-7}$, $l=e,\mu,\tau$,
        at the $Z$-pole.
        The dashed red (solid blue) curve corresponds to the LNC
        (LNC + LNV) contribution.
        The green dotted line corresponds to the case of $M_2 - M_1 = 0$ GeV.
        See text for further details.
    }
    \label{fig:YukawaBreakingZpoleM40}
\end{figure}

 There is a third possibility, namely, that the
  LNV contribution is smaller than the LNC contribution, but is not
  negligible. In this case both the forward-backward asymmetries
  generated by the sum of the LNV and LNC contributions, that would typically
  correspond to the production of a Majorana particle, $A^{M}_{FB}$, and
  that generated only by the
  LNC contribution corresponding to production of a Dirac fermion,
  $A^{D}_{FB}$, would be non-zero: $A^{M}_{FB}\neq 0$, $A^{D}_{FB}\neq 0$.
  However, due to the LNV contribution we would have
  $A^{M}_{FB} \neq A^{D}_{FB}$ and $A^{M}_{FB} < A^{D}_{FB}$.
  This is illustrated in Fig. \ref{fig:YukawaBreakingZpoleM40}, showing
  the differential cross sections $d\sigma_{\rm M,D}/d\cos\theta$
  of  production of $N_{1,2}$ with  $M_1 \cong M_2 = 40$ GeV,
  $M_2-M_1 \cong 0.0414$ eV and
   $((RV)_{l1,2})^2 = 2.5\times10^{-7}$, $l=e,\mu,\tau$,
  at the $Z$-pole. In this case, as can be shown, we get
  at  $q^2_N = M^2_{1,2}$:
  $|P^{Z}_{\mathrm{LNC},2}/P^Z_{\mathrm{LNV},2}|^2 \cong |1 + i \Gamma_{1,2}/(M_2-M_1)|^2
  \cong 1.72$. 
  The asymmetry 
  $A^{D}_{FB}$ is by a factor of 2.4 bigger than
  the potentially measurable $A^{M}_{FB}$:
  $A^{D}_{FB}\cong 0.144$, $A^{M}_{FB}\cong 0.060$.
  Thus, if it were possible to calculate
  $A^{D}_{FB}$, the difference with the observed 
  asymmetry would be an indication that there is a
  LNV contribution to the asymmetry and that one (or more than one but
  with directly unobservable mass splittings) 
  Majorana particle (or particles) is (are) produced.
  The calculation of  $A^{D}_{FB}$ requires the knowledge of
  the mass  $M_N\cong M_{1,2}$   and the couplings of
  $(RV)_{lN} =\sqrt{2} (RV)_{l2}$ of
  the hypothetical Dirac fermion $N$. These potentially can be determined
  from the observation of the Breit-Wigner resonance enhancement
  of the cross section 
  $d\sigma(e^+ + e^- \rightarrow \mu^- + j_1 + j_2 + E_{mis})/dM_{\mu^-j_1j_2}$
  at $M_{\mu^-j_1j_2} \cong M_N = M_1 \cong M_2$. We remark finally that
  in the considered example of $M_{1,2} \cong 40$ GeV,
  the mean free path of $N_{1,2}$ is $d_{1,2}\cong 5.18\times 10^{-4}$ cm. 
%

We note that with the considered couplings, approximately
  $4.7\times 10^{5}$,  $1.2\times 10^{6}$ and $8.6\times 10^{5}$
  $N_{1,2}$ with the discussed masses respectively of
  $M_{1,2} \cong 60$, 10 and 40 GeV are predicted
  to be produced at FCCee during its first phase
  after four years of running.

Thus, we see that in the considered seesaw scenario in which
the splitting of the masses of $N_{1,2}$ is given by
$M_2 - M_1 = m_3 - m_2\cong 0.0414$ eV,
depending on the values of the  masses of $N_{1,2}$ and of 
their charged (neutral) current couplings $(RV)_{l1,2}$,
the production of $N_{1,2}$ at the $Z$-pole
can mimic either the production of a single Dirac fermion or the
production of a single Majorana particle.
In order to understand what actually is the nature
of the fermion $\bar{N}$ produced and how many fermions are ``hidden''
by the production of the single one, it is of crucial
importance to measure the couplings $(RV)_{l\bar{N}}$ of $\bar{N}$.

Suppose that in the considered process 
$e^+ + e^- \rightarrow \mu^- + j_1 + j_2 + E_{mis}$ the production of a single
fermion $N$ with mass $M_N$ and CC (NC) couplings $(RV)_{lN}$
is observed. Interpreting these data in terms of a symmetry-protected
seesaw scenario with two RH neutrinos, and thus, two heavy Majorana neutrinos
$N_{1,2}$ with masses $M_{1,2}$, would lead to the conclusion that
$M_1 \cong M_2 = M_N$, $|(RV)_{l1}| \cong |(RV)_{l2}| =|(RV)_{lN}|/\sqrt{2}$,
with $|M_2-M_1| \ll M_{1,2}$ such that the two masses $M_{1,2}$ cannot be resolved
experimentally, and $\Gamma_1 \cong \Gamma_2$.
Knowing $M_{1,2}$ and $(RV)_{l1,2}$, one can determine  $\Gamma_{1,2}$.
If in the production of $N$ one finds $A_{FB} \neq 0$, that would imply that
$|M_2-M_1|^2 \ll (<) \Gamma^2_{1,2}$. In contrast, a zero forward-backward asymmetry,
$A_{FB} =  0$, would mean that $|M_2-M_1|^2 \gg \Gamma^2_{1,2}$.
In this case an upper limit on $|M_2-M_1|$ can be obtained from the
non-observability of each of the two heavy Majorana neutrinos $N_{1,2}$
due to insufficiently good  $M_{l^-j_1j_2}$-resolution, which does not allow one
to resolve the masses $M_1$ and $M_2$:
 $|M_2-M_1| < (\ll) \delta M_{l^-j_1j_2}$. 

The preceding considerations apply also in the cases when the
contributions of the virtual $W^\pm$-exchange diagrams in the
process of interest have to be taken into account or are the dominant ones.

\vspace{0.3cm}
{\bf Scenario with $L^\prime$-symmetry  breaking by the Yukawa coupling and the
  RH neutrino Majorana mass term}

\vspace{0.3cm}
In this scenario with two heavy Majorana neutrinos $N_{1,2}$
discussed in subsection 
\ref{sec:LpB3}, both their 
mass difference $(M_2-M_1)$ and their
couplings $(RV)_{li}$ can be relatively large, which opens up the possibility
to both observe $N_{1,2}$ and determine their Majorana nature.
We consider the production of $N_{1,2}$ at the $Z$-pole
in the following three cases:
i) $M_{12} = 10.0$ GeV, $M_{22} = 1.0$ GeV,
ii) $M_{12} = 40.0$ GeV, $M_{22} = 4.0$ GeV, and 
iii)  $M_{12} = 40.0$ GeV, $M_{22} = 2.0$ GeV.
In all these cases $\rho_1 = -\,1$ and $\rho_2 = 1$.
For the study which follows  we set
$m^D_{l 2} = 10^{-3}$ GeV, $l=e,\mu,\tau$.
We recall that to a very good approximation
$M_{1,2} = 0.5[\sqrt{M^2_{22} + 4M^2_{12}} \mp M_{22}]$, 
$(RV)_{l1} = m^D_{l2}\sin\varphi/M_1$,
$(RV)_{l2} = m^D_{l2}\cos\varphi/M_2$,
$\sin\varphi = \sqrt{M_1}/\sqrt{M_1+M_2}$.
\begin{figure}[t]
    \centering
 \includegraphics[width=0.96\linewidth]{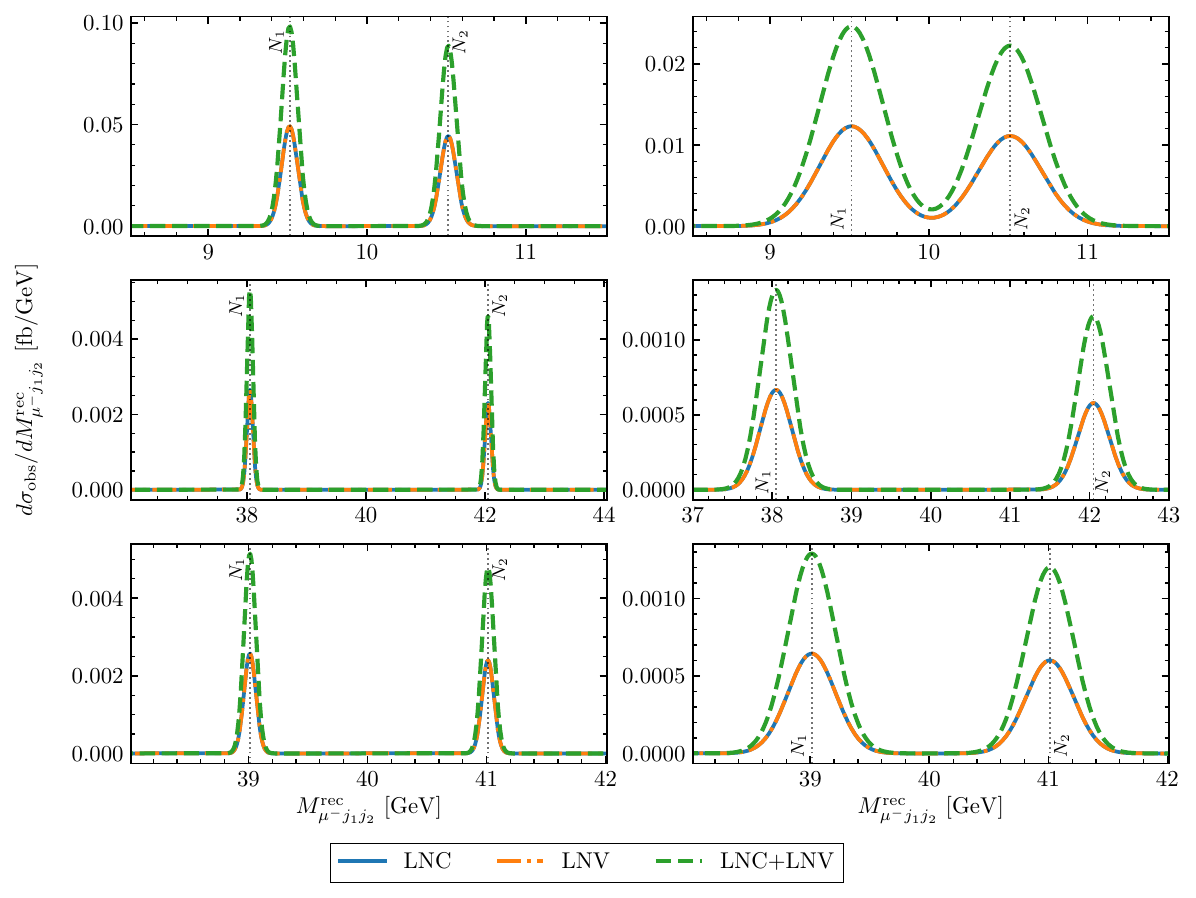}
 \caption{The differential cross sections
     $d\sigma(e^+ + e^- \rightarrow \mu^- + j_1 + j_2 + E_{mis})/dM_{\mu^-j_1j_2}$
      at the $Z$-pole,
      $\sqrt{s} = M_Z = 91.1876$ GeV, for 
      $M_1 = 9.513$, 38.050 and 39.013 GeV and respectively 
      $M_2 = 10.513$, 42.050 and 41.013 GeV, assuming
      the measured $M_{\mu^-j_1j_2}$ follows the Gaussian distribution with
      a $1\sigma$ error of $\delta M_{l^-j_1j_2} =  0.05$ GeV and 0.20 GeV
       and taking $m^D_{\ell 2} = 10^{-3}$ GeV.
See text for further details.}
    \label{fig:2RHZpole1Gauss}
\end{figure}
%
\noindent Using these ingredients we find:
\begin{eqnarray}
    \nonumber
    P^Z_{\mathrm{LNV},2} = \dfrac{m^D_{l2}\sum_{l'=e,\mu,\tau} m^D_{l'2}}{\sqrt{q^2_N}}\,
    \dfrac{M_2 - M_1 - i \dfrac{\Gamma_2 M_2 - \Gamma_1 M_1}{M_2 + M_1}}
    {[q_N^2-M_2^2+iM_2\Gamma_2][q_N^2-M_1^2+iM_1\Gamma_1]}\,. 
\label{eq:P2ZLNVLbYMR}           
\end{eqnarray}
%
Thus, $P^{Z}_{\mathrm{LNV},2} \propto (M_2 - M_1)$
\footnote{We also have $P^{W}_{\mathrm{LNV},2} \propto (M_2 - M_1)$.}
and since $(M_2 - M_1) = 1$, 2
and 4 GeV in the considered cases, the LNV contribution to the
cross section of the process of interest seems to be 
rather suppressed. However, in a similar fashion we get:
\begin{eqnarray}
    \nonumber
    P^Z_{\mathrm{LNC},2} = \dfrac{m^D_{l2}\sum_{l'=e,\mu,\tau} m^D_{l'2}}{M_1M_2}\,
    \dfrac{q^2_N -\,\dfrac{M^3_2 + M^3_1}{ M_2+ M_1}
      + i \dfrac{\Gamma_2 M^2_2 + \Gamma_1 M^2_1}{M_2 + M_1}}
    {[q_N^2-M_2^2+iM_2\Gamma_2][q_N^2-M_1^2+iM_1\Gamma_1]}\,. 
\label{eq:P2ZLNCLbYMR}         
\end{eqnarray}
%
For the ratio  $|P^Z_{\mathrm{LNV},2}|^2/|P^Z_{\mathrm{LNC},2}|^2$ at
$q^2_N \equiv M^2_{\mu^-j_1j_2} = M^2_{2(1)}$ we obtain taking into account that
in the considered case
of $N_2 - N_1$ mass splitting 
$|\Gamma_2 M_2 - \Gamma_1 M_1|/(M^2_2 - M^2_1) \ll 1$
and $(\Gamma_2 M^2_2 + \Gamma_1 M^2_1)/(M^2_{2(1)}(M_2 + M_1)) \ll 1$:
\begin{equation}
  \dfrac{|P^Z_{\mathrm{LNV},2}|^2}{|P^Z_{\mathrm{LNC},2}|^2} =
  \dfrac{|\sum_{j=1}^{2} F^{Z}_{{\rm LNV},j}(M^2_{2(1)})|^2}
        {|\sum_{j=1}^{2} F^{Z}_{{\rm LNC},j}(M^2_{2(1)})|^2} = 1\,.
        \label{eq:PZLNVoPZLNC}
\end{equation}
%

\begin{figure}[t]
    \centering
 \includegraphics[width=0.99\linewidth]{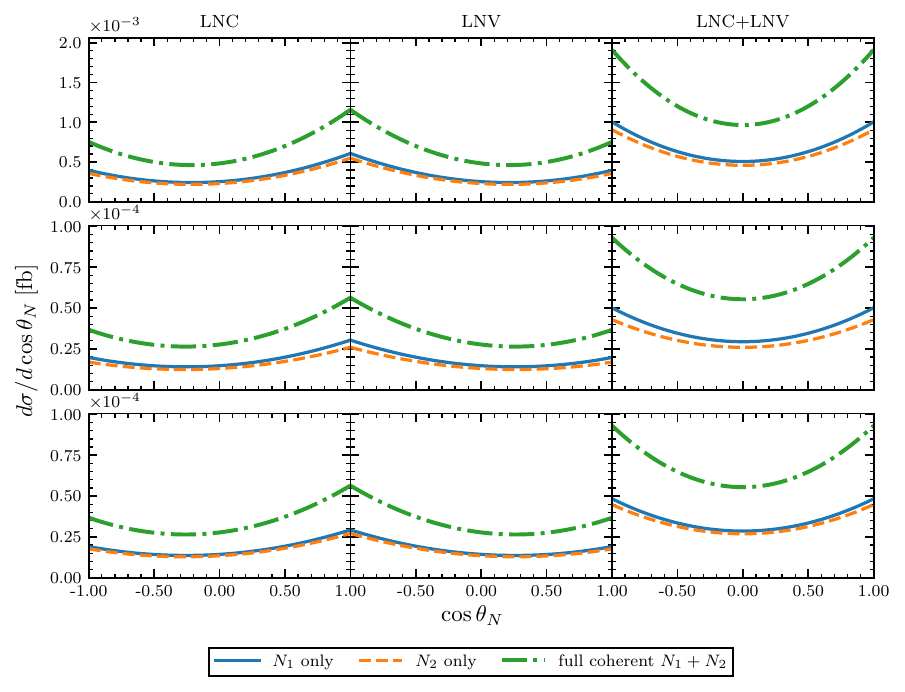}
  \caption{The differential cross sections
        $d\sigma_{\rm LNC}(|A_W + A_Z|^2)/d\cos\theta$ (left panel), $d\sigma_{\rm LNV}(|A_W + A_Z|^2)/d\cos\theta$ (middle panel), and 
        $d\sigma_{\rm M,D}/d\cos\theta$ (right panel)
        at the $Z$-pole,
        $\sqrt{s}=M_Z=91.1876$ GeV.
        The top, middle and bottom rows correspond, respectively, to
        $(M_1,M_2)=(9.513,10.513)$ GeV,
        $(38.050,42.050)$ GeV and
        $(39.013,41.013)$ GeV.
        See text for further details.
    }
    \label{fig:2RHZpole2}
\end{figure}
%

\noindent This result implies that in  the range of
$q^2_N$ of interest, 
$d\sigma_{\rm LNV}(|A_W + A_Z|^2;\cos\theta)/d\cos\theta \cong 
d\sigma_{\rm LNC}(|A_W + A_Z|^2;-\cos\theta)/d\cos\theta$
\footnote{At the $Z$-pole the $W$-exchange diagrams give  negligible
contributions and $d\sigma_{\rm LNV,LNC}(|A_W + A_Z|^2)/d\cos\theta = 
d\sigma_{\rm LNV,LNC}(|A_Z|^2)/d\cos\theta$.}
and therefore $d\sigma_{\rm M}/d\cos\theta$
will exhibit the  F-B symmetry signature, i.e., $A_{FB} = 0$,
of production of heavy Majorana neutrinos. 

We note  that for $M^2_j \ll s$, the $\cos\theta$-dependence of the
differential cross sections $d\sigma_{\rm LNC,LNV}(|A_W + A_Z|^2)/d\cos\theta$ and
$d\sigma_{\rm M,D}/d\cos\theta$, as it follows
from Eqs. (\ref{eq:WdiffLNC}) - (\ref{eq:ZLNV}), is not sensitive to
$M_j$, while their magnitudes depend on $M_j$.
For $s = M^2_Z$ and the cases i) - iii) we are considering,
the effect of the masses of $N_{1,2}$ on the shape of
$d\sigma_{\rm M,D}/d\cos\theta$ as a function of $\cos\theta$
is rather  weak and therefore hardly observable.
Thus, the $\cos\theta$-dependence of, e.g., $d\sigma_{\rm M}/d\cos\theta$,
integrated over $q^2_N$, 
will mimic that of production of a single heavy Majorana neutrino.

To establish the existence of two heavy Majorana neutrinos $N_{1,2}$ in
the discussed seesaw scenario 
with $M^2_{1,2} \ll s$  one has to study experimentally
the $q^2_N$-dependence of the cross section of the process 
$e^+ + e^- \rightarrow l^- + j_1 + j_2 + E_{mis}$,
where $q^2_N = (p(l^-) + p(j_1) + p(j_2))^2 \equiv M^2_{l^-j_1j_2}$ and 
$E_{mis}$ stands for ``missing energy'', $l^-=e^-,\mu^-$,
and search
for Breit-Wigner resonance enhancements that should be present at
$M_{l^-j_1j_2} = M_1$ and at $M_{l^-j_1j_2} = M_2$.
The total widths of $N_{1,2}$, $\Gamma_{1,2}$,  
which determine the widths of the two resonances, are quite small:
in the case of $M_{1,2} < M_W$ we have
$\Gamma_{1,2}\propto G^2_FM^5_{1,2} \sum_l |(RV)_{l1,2}|^2$, while
if $M_{1,2} >  M_W$, $\Gamma_{1,2}\propto G_FM^3_{1,2}\sum_l |(RV)_{l1,2}|^2$.
In addition the mass difference $(M_2 - M_1)$ can be relatively small.
This implies that the search for resonance enhancements of
$d\sigma(e^+ + e^- \rightarrow l^- + j_1 + j_2 + E_{mis})/dM_{l^-j_1j_2}$
should be performed with sufficiently small steps in $M_{l^-j_1j_2}$,
$\Delta M_{l^-j_1j_2}$, as small as experimentally possible.

We show in Fig. \ref{fig:2RHZpole1Gauss} the results of such a search at the
$Z$-pole with $\Delta M_{l^-j_1j_2} =  0.05$ GeV and 0.20 GeV
in the three cases of masses of $N_{1,2}$ specified at the beginning
of the present subsection and for $m^D_{l 2} = 10^{-3}$ GeV, $l=e,\mu,\tau$.
The plots shown in
Fig. \ref{fig:2RHZpole1Gauss} are obtained 
by taking into account the uncertainty in the measurement of
$M_{l^-j_1j_2}$ assuming that the measured $M_{l^-j_1j_2}$
follows the Gaussian distribution with $1\sigma$ error 
of $\delta M_{l^-j_1j_2} =  0.05$ GeV and 0.20 GeV.
The resonance enhancement is clearly present at
$M_{l^-j_1j_2} = M_1$ and at $M_{l^-j_1j_2} = M_2$.

We note that for the parameters used the mean free paths of
$N_1$ and $N_2$ with masses $M_1 = 9.51$ GeV and $M_2 = 10.51$ GeV
are approximately 1.5 m and 0.9 m, respectively. In the other two cases
of $N_{1,2}$ with masses $M_1 = 38.05$ GeV, $M_2 = 42.05$ GeV,
and $M_1 = 39.013$ GeV, $M_2 = 41.013$ GeV,
the mean free paths are,
respectively,  $5.0\times 10^{-3}$ m, $2.9\times 10^{-3}$ m, and 
$4.4\times 10^{-3}$ m, $3.3\times 10^{-3}$ m.
This would lead to the well known ``displaced decay''
signature of production of $N_{1,2}$ with practically zero
background.

If a resonance enhancement of
$d\sigma(e^+ + e^- \to l^- + j_1 + j_2 + E_{mis})/dM_{l^-j_1j_2}$
at $M_{l^-j_1j_2} \cong M_1$ and $M_{l^-j_1j_2} \cong M_2$ is observed, 
that would signal the production of two fermions $N_{1,2}$ with
masses $M_{1,2}$, but no indications of their nature - Dirac or Majorana -
will be obtained. The nature of $N_{1,2}$ can be tested by studying
the $\cos\theta$ dependence of
$d^2\sigma(e^+ + e^- \to l^-+j_1+j_2+E_{mis})/(dM_{l^-j_1j_2}d\cos\theta)$
at $M_{l^-j_1j_2} \cong M_{1}$ and  $M_{l^-j_1j_2} \cong M_{2}$.
For Majorana $N_{1,2}$ we would have $A_{FB} = 0$ at the two rather narrow
``windows'' $M_{l^-j_1j_2} \cong M_{1,2}$ with two distinct
$\cos\theta$-distributions,
$d^2\sigma(e^++e^-\to l^-+j_1+j_2+E_{mis};M_{l^-j_1j_2}=M_1)/(dM_{l^-j_1j_2}d\cos\theta)\neq d^2\sigma(e^++e^-\to l^-+j_1+j_2+E_{mis};M_{l^-j_1j_2}=M_2)/(dM_{l^-j_1j_2}d\cos\theta)$.
If the masses $M_1$ and $M_2$ cannot be resolved experimentally, 
the $\cos\theta$-dependence of the
cross section
$d\sigma(e^+ + e^- \to l^-+j_1+j_2+E_{mis})/d\cos\theta$ would  mimic that of
production of a single Majorana fermion with  $A_{FB} = 0$.

These features of
$d^2\sigma(e^++e^-\to l^-+j_1+j_2+E_{mis};M_{l^-j_1j_2}=M_1)/(dM_{l^-j_1j_2}d\cos\theta)$ and
$ d^2\sigma(e^++e^-\to l^-+j_1+j_2+E_{mis};M_{l^-j_1j_2}=M_2)/(dM_{l^-j_1j_2}d\cos\theta)$ are illustrated in Fig.  \ref{fig:2RHZpole2}.
As Fig.  \ref{fig:2RHZpole2}
shows, the difference between the $\cos\theta$-distributions
at $M_{l^-j_1j_2} \cong M_1$ and $M_{l^-j_1j_2} \cong M_2$ is rather small
and experimentally the two distributions might be indistinguishable.
However, even the presence of an apparently identical
$\cos\theta$-distribution with $A_{FB} = 0$ at two different values of
$M_{l^-j_1j_2}$, namely, at $M_{l^-j_1j_2} \cong M_1$ and $M_{l^-j_1j_2} \cong M_2$,
would be evidence of production of two heavy Majorana neutrinos
with different but close masses. 

The tests discussed above can 
  be performed at FCCee running at the $Z^0$-pole.
  The estimated numbers of $N_{1,2}$ expected to be produced in
  four years of data taking in the considered cases of masses of 
  $(M_1,M_2) = (9.513,10.513)$, (38.050,42.050) and (39.013,41.013) GeV
  read, respectively, $(1.26\times 10^4,1.13\times 10^4)$,
  (591,495) and (567,519). 

\begin{figure}[t]
    \centering
    \includegraphics[width=0.80\linewidth]{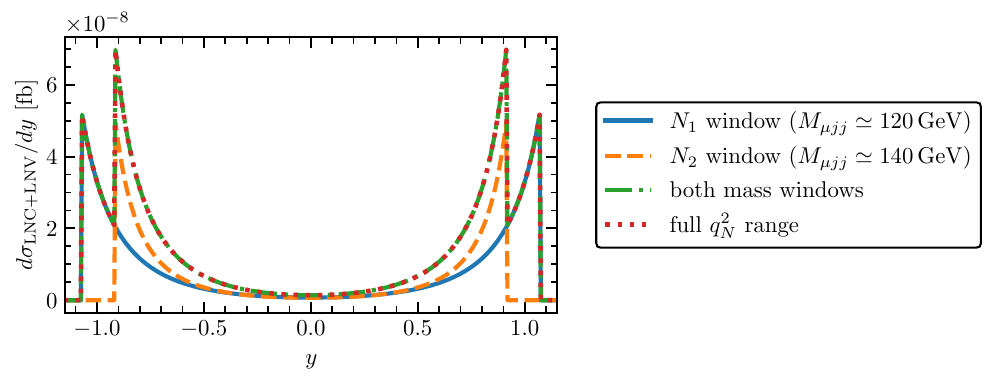} 
    \caption{The differential cross sections
      $d\sigma(e^+ + e^- \rightarrow l^- + j_1 + j_2 + E_{mis})/dy$
      (rapidity distribution) at $\sqrt{s} = 350$ GeV for 
      $M_1 = 120$ GeV and 
      $M_2 = 140$ GeV.
      See text for further details.}
    \label{fig:2RH350y}
\end{figure}

In the cases when $M^2_{1,2}/s$ is not negligible and $\sqrt{s} > 160$ GeV,
where the contributions of the $Z$-boson exchange diagrams are insignificant and
$d\sigma(e^++e^-\to l^-+j_1+j_2+E_{mis})/dM_{l^-j_1j_2}$ is determined by the
$W^-$-exchange diagrams, the presence of the two heavy Majorana neutrinos
$N_{1,2}$ can be inferred by studying the rapidity distribution 
$d\sigma(e^++e^-\to l^-+j_1+j_2+E_{mis})/dy$, where the rapidity
\begin{equation}
  y = \dfrac{1}{2}\, \ln \dfrac{E_{l^-j1j2} + |{\bf k}_{l^-j1j2}|\cos\theta}
  {E_{l^-j1j2} - |{\bf k}_{l^-j1j2}|\cos\theta}\,.
  \label{eq:yN12}
\end{equation}
%
 Here $E_{l^-j1j2}$ and ${\bf k}_{l^-j1j2}$
are the sums of the energies and of the 3-momenta of $l^-$, $j_1$ and $j_2$.
This distribution, as we have discussed, will be symmetric with respect to
$y=0$ and will have two distinct maxima at $y < 0$ located at
$M_{l^-j_1j_2} = M_{1,2}$, which will be a clear evidence that the produced
$N_{1,2}$ are Majorana particles. The observation of the two maxima
requires a sufficiently high precision determination of the rapidity $y$
over the wide range of its kinematically allowed values. The discussed
possibility is illustrated in Fig. \ref{fig:2RH350y}.

%
\subsubsection{Seesaw Scenario with Three RH Neutrinos}
\label{sec:3RHAFB}
%

This seesaw scenario with three heavy Majorana neutrinos
$N_{1,2,3}$ with masses $M_1 = 9$ GeV, $M_2 = 11$ GeV and $M_3=10$ GeV
is discussed in subsection \ref{sec:3RH}, where the $N_{1,2,3}$
couplings $(RV)_{lj}$, $l=e,\mu,\tau$, $j=1,2,3$,
are also given. 
\begin{figure}[t]
    \centering
 \includegraphics[width=0.96\linewidth]{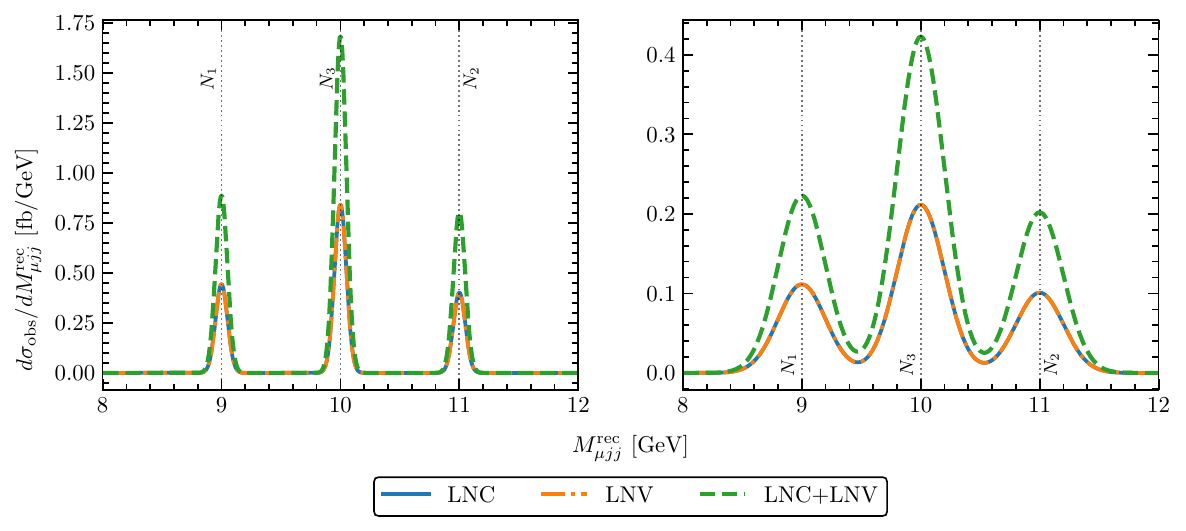}
 \caption{The differential cross sections
     $d\sigma(e^+ + e^- \rightarrow \mu^- + j_1 + j_2 + E_{mis})/dM_{\mu^-j_1j_2}$
      at the $Z$-pole,
      $\sqrt{s} = M_Z = 91.1876$ GeV, in the scenario with three RH neutrinos
      with heavy Majorana neutrino masses 
      $M_1 = 9$ GeV, $M_2 = 11$ GeV and $M_3=10$ GeV assuming
      the measured $M_{\mu^-j_1j_2}$ follows the Gaussian distribution with
      a $1\sigma$ error of $\delta M_{\mu^-j_1j_2} =  0.05$ GeV and 0.20 GeV.
See text for further details.}
    \label{fig:3RHZpoleGauss}
\end{figure}
%
\begin{figure}[t]
    \centering
 \includegraphics[width=0.99\linewidth]{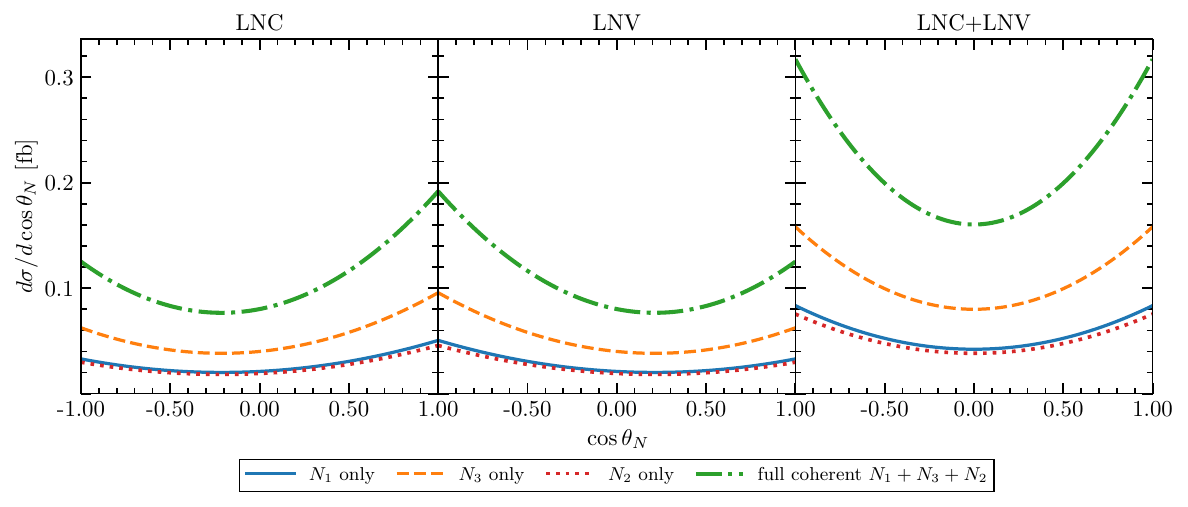}
    \caption{The differential cross sections
        $d\sigma_{\rm LNC}(|A_W + A_Z|^2)/d\cos\theta$ (left panel), $d\sigma_{\rm LNV}(|A_W + A_Z|^2)/d\cos\theta$ (middle panel), and 
        $d\sigma_{\rm M,D}/d\cos\theta$ (right panel)
    of the process
    $e^+ + e^- \to \mu^- + j_1 + j_2 + E_{mis}$ at the $Z$-pole,
    $\sqrt{s} = M_Z = 91.1876$ GeV, in the seesaw scenario with three RH and three
    heavy Majorana neutrinos $N_{1,2,3}$ with masses
    $M_1 = 9$ GeV,
    $M_2 = 11$ GeV and $M_3=10$ GeV.
    See text for further details.}
    \label{fig:3RHZpole}
\end{figure}
%
We show in Fig. \ref{fig:3RHZpoleGauss} the cross section 
$d\sigma(e^+ + e^- \to \mu^- + j_1 + j_2 + E_{mis})/dM_{\mu^-j_1j_2}$
at the $Z$-pole, $\sqrt{s} = M_Z$, in the considered seesaw scenario. 
The plots shown in Fig. \ref{fig:3RHZpoleGauss} are obtained
by taking into account the uncertainty in the measurement of
$M_{\mu^-j_1j_2}$ assuming that the measured $M_{\mu^-j_1j_2}$
follows the Gaussian distribution with $1\sigma$ error 
of $\delta M_{\mu^-j_1j_2} =  0.05$ GeV and 0.20 GeV.
The resonance enhancement is clearly present at
$M_{\mu^-j_1j_2} = M_1$, $M_{\mu^-j_1j_2} = M_2$ and  $M_{\mu^-j_1j_2} = M_3$.

The $\cos\theta$-dependence of the cross sections
    $d\sigma_{\rm LNC,LNV}(|A_W + A_Z|^2)/d\cos\theta$  and 
    $d\sigma_{\rm M}/d\cos\theta$ 
of the process $e^+ + e^- \to \mu^- + j_1 + j_2 + E_{mis}$ with
$\sqrt{s} = M_Z$ ($Z$-pole)
and for $M_{\mu^-j_1j_2} = M_1$, $M_{\mu^-j_1j_2} = M_2$ and  $M_{\mu^-j_1j_2} = M_3$  
is presented in Fig. \ref{fig:3RHZpole}.
We recall that the case denoted  by LNC+LNV (LNC) corresponds to
the production of a Majorana (Dirac) particle.
In all three cases of LNC+LNV associated with production of
$N_{1,2,3}$ ($M_{\mu^-j_1j_2} = M_{1,2,3}$)  
we find $A_{FB} = 0$, which is a clear signature that $N_{1,2,3}$ are
Majorana fermions. As Fig. \ref{fig:3RHZpole} indicates,
the difference between $d\sigma_{\rm M}/d\cos\theta$
at $M_{\mu^-j_1j_2} = M_{1}$ and $M_{\mu^-j_1j_2} = M_{2}$
is hardly observable, but the fact that $A_{FB} = 0$
at two different values of $M_{\mu^-j_1j_2}$ would be evidence, if not
proof, of the production of two  heavy Majorana neutrinos with
just different masses. At the same time 
$d\sigma_{\rm M}/d\cos\theta$ at $M_{\mu^-j_1j_2} = M_{3}$
differs noticeably from $d\sigma_{\rm M}/d\cos\theta$ at
$M_{\mu^-j_1j_2} = M_{1},M_2$.

For the values of the masses $M_{1,2,3}$ and
couplings  $(RV)_{lj}$ used, the mean free paths of
$N_1$, $N_2$ and $N_3$ read, respectively,
0.68 m, 0.19 m and 0.22 m. Thus,  they all will decay within the detector
with the ``displaced decay'' signature of $N_{1,2,3}$ production with
practically zero background.

The F-B symmetry test of the Majorana nature of the heavy neutrinos
  $N_{1,2,3}$ discussed here
  can be  performed at FCCee running at the $Z^0$-pole, 
  where in four years a total number of produced
  $N_{1,2,3}$, summed over the three flavour couplings of each of them,
  is estimated to reach $1.6 \times 10^5$.

We note finally that if $N_{1,2,3}$ were Dirac fermions,  only the LNC
diagrams would contribute to the cross section of interest.
In this case the forward-backward asymmetry
$A_{FB}$ would be non-zero at $M_{\mu^-j_1j_2} = M_{1,2,3}$ and would
have the values $A_{FB} = 0.157$,  0.156,  0.157, respectively.

%
\section{Summary and Conclusions}
\label{sec:SC}
%

In the present article we have investigated the  possibility to determine
the Majorana nature of the HNLs, which are an integral part of the
minimal type I seesaw  mechanism of light neutrino mass generation,
in experiments at $e^+ - e^-$ (FCCee, CEPC, ILC, CLIC) and
$\mu^+ - \mu^-$ colliders. This was done on the examples of
two low-scale seesaw scenarios with two RH neutrinos and one
low-scale scenario with three RH neutrinos.
The scenarios with two RH neutrinos are of the ``symmetry-protected'' class,
in which the respective heavy Majorana neutrinos $N_{1,2}$ form a pseudo-Dirac
pair. These scenarios rely on a small explicit
breaking of a $U(1)_{L'}$ symmetry, associated with a conserved
non-standard lepton charge $L'$.
In both considered cases  $N_{1,2}$ have relatively large charged (neutral)
current  weak interaction couplings with the charged leptons and
the SM $W^\pm$-bosons (flavour neutrinos and the $Z^0$-boson).
The two scenarios differ by the way the  $U(1)_{L'}$ symmetry
is broken: in the first scenario the breaking originates from
the neutrino Yukawa coupling, while in the second it is generated
by both the  neutrino Yukawa coupling and the RH neutrino Majorana mass
term. In the first scenario the splitting $|M_2 - M_1|$ is determined
by the difference of the two non-zero light neutrino masses and thus
is exceedingly small. We showed that in the
second scenario there is a region of the
respective parameter space where the $N_{1} - N_{2}$ mass splitting,
$|M_2-M_1|$, can be sizeable. As a consequence, the phenomenologically
allowed magnitude of the mass difference $|M_2-M_1|$,
which in the considered scenarios is a lepton charge violating quantity
directly related to the Majorana nature of $N_{1,2}$, can 
differ drastically in the two cases. 
The seesaw scenario with three RH neutrinos
is one example of seesaw models investigated recently in
\cite{Granelli:2026qss} in which both the charged (neutral) couplings of the
heavy Majorana neutrinos $N_{1,2,3}$ and the differences between their masses
can be relatively large and might allow their observation
in experiments at $e^+ - e^-$ ($\mu^+ - \mu^-$) colliders.

After analysing in some detail the considered specific seesaw
realisations, we discuss next the production of the heavy Majorana neutrinos
$N_{j}$, $j=1,...,k$, $k=2$ or 3, present in these realisations,
in $e^+ - e^-$ ($\mu^+ - \mu^-$) collisions.
Each heavy Majorana neutrino $N_{j}$ can be produced in the process of 
$e^+ + e^-$ ($\mu^+ + \mu^-$) annihilation together with
$\bar{\nu}_e$ ($\bar{\nu}_\mu$) or $\nu_e$ ($\nu_\mu$) via virtual 
$W^+$ or $W^-$ exchange diagrams, and with $\bar{\nu}_{\ell'}$ or
$\nu_{\ell'}$, $\ell' = e,\mu,\tau$, via $Z^0$ exchange diagrams
(Fig. \ref{fig:HNL_production_decay}). 
We assume that $N_j$ undergoes the decay  $N_j \to l^- + W^+$,
$l=e,\mu,\tau$, with virtual or real  $W^+$ producing two identifiable
jets in the final state via $W^+ \to j_1 + j_2$
\footnote{The case of $N_j \to l^+ + W^-$ decay can be treated similarly
to that considered by us and leads to the same results.}.
When $N_j$ is produced with $\bar{\nu}_e$ or  $\bar{\nu}_{\ell'}$,
the decay $N_j \to l^- + W^+$ is allowed independently of whether
$N_j$ is a Majorana or Dirac fermion since the corresponding chain of
processes is Lepton Number Conserving (LNC).
However, in the case of $e^+ + e^- \to N_j + \nu_e (\nu_{\ell'})$,
the same decay $N_j \to l^- + W^+$ is allowed only if $N_j$ is a Majorana
particle and is forbidden for a Dirac $N_j$ as the corresponding chain
of processes is Lepton Number Violating (LNV).
The presence of the LNV contribution in addition to the LNC one
in the differential  cross section of the process of interest,
$e^+ + e^- \to N_j + \overline{\nu_{\ell'}}~(\nu_{\ell'}) \to l^- + W^+ + E_{mis} \to l^- + j_1 + j_2 + E_{mis}$,
$d\sigma(e^+ + e^-\to l^- + j_1 + j_2 + E_{mis})/d\cos\theta$,
where $E_{mis}$ stands for missing energy,
$(p(l^-)+p(j_1)+p(j_2))^2 \equiv M^2_{l^-j_1j_2} = M^2_j$,
and $\theta$ is the angle between the 3-momenta of $N_j$ and $e^-$
in the $e^+ - e^-$ c.m.s., $p(l^-)$, $p(j_1)$ and $p(j_2)$ being the
4-momenta respectively of $l^-$, $j_1$ and $j_2$,
ensures that the $\cos\theta$-distribution of the produced $N_j$ is
fully forward-backward (F-B) symmetric. The zero F-B asymmetry,
$A_{FB} = 0$, in the production of $N_j$, thus, would be
a signature of the Majorana nature of $N_j$;
in the absence of the LNV contribution,
the  $\cos\theta$-distribution of interest generated by the LNC contribution
would exhibit a non-zero  F-B asymmetry, $A_{FB} \neq 0$, 
which would  signal the production of a Dirac fermion
\cite{Petcov:1984nf}.

Using the forward-backward symmetry/asymmetry indicators we proceeded
to investigate the
possibility to determine the Majorana nature of the heavy Majorana neutrinos
$N_r$ present in the considered seesaw scenarios.
We derived analytical expressions for the LNC and LNV contributions
in the cross section
$d\sigma(e^+ + e^-\to l^- + j_1 + j_2 + E_{mis})/d\cos\theta$
when the $W^+$-boson of the decay $N_j \to l^- + W^+$ is real
and either the virtual $W^\pm$- or $Z^0$- exchange diagrams give the dominant
contribution. The derived analytical results allowed us to identify
simple factors in $d\sigma(e^+ + e^-\to l^- + j_1 + j_2 + E_{mis})/d\cos\theta$
that permitted to obtain a quantitative evaluation of the relative magnitude of
the LNV contribution, which is crucial to our analysis, in the differential
cross section. In our numerical analysis we took into account
the interference between the contributions to the amplitude of the
process of interest of the $W^\pm$-exchange and $Z^0$-exchange diagrams
whenever it was non-negligible.
We presented results for the considered three seesaw scenarios
in the cases of production of $N_r$ with masses
$10~{\rm GeV} \lesssim M_j < M_Z$, $M_Z \cong 91$ GeV
being the $Z^0$-boson mass, in $e^+ + e^-$ collisions at the $Z^0$-pole,
$\sqrt{s} = M_Z$, as well as in the case of $\sqrt{s} = 350$ GeV and masses
of $N_{1,2}$ of one of the scenarios
with two RH neutrinos, in which the splitting  $|M_2 - M_1|$
can be relatively large,
of $M_1 = 120$ GeV and $M_2=140$ GeV. These results were obtained
without accounting for the prospective experimental conditions
under which the discussed experiments would be performed:
evaluating the feasibility
of the considered observation and determination of the Majorana nature of the
heavy neutrinos in the discussed seesaw scenarios was beyond the scope of
our study and will be done elsewhere. We illustrated graphically the
major aspects of our analysis and the results we have obtained.

We found that depending on the magnitude of the mass splittings
$|M_i - M_j|$, $i\neq j$,
the total decay widths $\Gamma_j$ of
the heavy Majorana neutrinos $N_j$, which are determined by the
mass $M_j$ and the charged and neutral current couplings
$(RV)_{lj}$ of $N_j$, as well as the precision
with which the observable $M_{l^-j_1j_2}$ can be determined
experimentally, the production of $N_{1,2}$ (or $N_{1,2,3}$)
at $e^+ - e^-$ (or $\mu^+ - \mu^-$) colliders
could mimic i) the production of a single Dirac fermion with
$A_{FB} \neq 0$, or ii) the production of a single Majorana fermion with
$A_{FB} = 0$. These possibilities can be realized
if the precision on $M_{l^-j_1j_2}$ is not sufficiently high
to allow one to resolve, e.g., the masses $M_1$ and $M_2$ of heavy
Majorana neutrinos $N_{1,2}$.
An example is the first seesaw scenario with two RH neutrinos considered
by us, in which the splitting $|M_2 - M_1|$ is determined by
the difference of the two non-zero light neutrino masses and thus
is exceedingly small (Figs. \ref{fig:YukawaBreakingZpole} and \ref{fig:2RHZpoleN12Gauss}).
In this case fixing $|(RV)_{l1,2}|$ to $5\times 10^{-4}$, $l=e,\mu,\tau$,
we have shown that if $M_{1} \cong M_2 = 60$ GeV,
the production of $N_{1,2}$ would mimic that of a single Dirac fermion
since one would observe a relatively large $A_{FB} \cong 0.127 \neq 0$,
while for $M_{1} \cong M_2 = 10$ GeV it would mimic
the production of a single Majorana particle as one would
find $A_{FB} \cong 0$.
The Breit-Wigner
resonance enhancement of the differential cross section
$d\sigma(e^+ + e^-\to l^- + j_1 + j_2 + E_{mis})/dM_{l^-j_1j_2}$
at $M_{l^-j_1j_2} \cong M_{1,2}$ would correspond in both cases
to a production of a single fermion.

If the data are compatible with the production of a single Dirac (Majorana)
fermion, interpreting the observations in terms of a symmetry-protected
seesaw scenario with two RH neutrinos, in which the corresponding
heavy Majorana neutrinos $N_{1,2}$ form a pseudo-Dirac pair,
one can derive an upper limit on $|M_2 - M_1|$ (if $A_{FB}\neq 0$)
or a range of possible values of $|M_2 - M_1|$ (when $A_{FB} = 0$),
provided it is possible
to determine $M_1\cong M_2$ and $|(RV)_{l1}| \cong |(RV)_{l2}|$ from the data.

In the second considered  symmetry-protected seesaw scenario
with sizeable $N_1-N_2$ mass splitting $|M_2 - M_1|$, 
the difference between the $\cos\theta$-
distributions at $M_{l^-j_1j_2} = M_1$ and at $M_{l^-j_1j_2} = M_2$
is rather small and experimentally the two distributions might be
indistinguishable. However, even the presence of an apparently identical
$\cos\theta$-distribution with $A_{FB} = 0$ at two different values of
$M_{l^-j_1j_2}$, namely, at $M_{l^-j_1j_2} \cong M_1$ and $M_{l^-j_1j_2} \cong M_2$,
would be  a proof of the production of two heavy
Majorana neutrinos with different but close masses. We showed also
that the production of  the two HNLs  can be observed via the Breit-Wigner
resonance enhancement of the differential cross section 
$d\sigma(e^+ + e^-\to l^- + j_1 + j_2 + E_{mis})/dM_{l^-j_1j_2}$
at $M_{l^-j_1j_2} \cong M_1$ and $M_{l^-j_1j_2} \cong M_2$
if the measured $M_{l^-j_1j_2}$ follows the Gaussian distribution
with $1\sigma$ error of $\delta M_{l^-j_1j_2} =  0.05$ GeV or 0.20 GeV
(Fig. \ref{fig:2RHZpole1Gauss}).

In the considered seesaw scenario with three RH neutrinos, the masses
of the heavy Majorana neutrinos
$N_{1,2,3}$ are fixed to $M_1 = 9$ GeV, $M_2 = 11$ GeV and $M_3=10$ GeV
and their charged (neutral) current  couplings are also fixed
(Eq. (\ref{eq:3NsRV})). We have considered the production of
$N_{1,2,3}$ in $e^+ - e^-$ collisions at the $Z^0$-pole. We showed that 
the presence of the three heavy Majorana neutrinos
can be revealed by the resonance enhancement of
the differential cross section 
$d\sigma(e^+ + e^- \to \mu^- + j_1 + j_2 + E_{mis})/dM_{\mu^-j_1j_2}$
at  $M_{\mu^-j_1j_2}\cong M_1$, $M_{\mu^-j_1j_2} \cong M_2$ and
$M_{\mu^-j_1j_2} \cong M_3$ (Fig. \ref{fig:3RHZpoleGauss}).
This result was obtained by taking into account the uncertainty in the
measurement of $M_{\mu^-j_1j_2}$, assuming that the measured $M_{\mu^-j_1j_2}$
follows the Gaussian distribution with $1\sigma$ error 
of $\delta M_{\mu^-j_1j_2} =  0.05$ GeV or 0.20 GeV.
The $\cos\theta$-distribution described by 
$d\sigma(e^+ + e^- \to \mu^- + j_1 + j_2 + E_{mis})/d\cos\theta$
was shown to exhibit forward-backward symmetry, i.e., $A_{FB} = 0$
at $M_{\mu^-j_1j_2} = M_{1,2,3}$ ( Fig. \ref{fig:3RHZpole}),
which is a clear signature that
the produced $N_{1,2,3}$ are Majorana fermions.
We found also that the difference between
the $\cos\theta$-distributions 
at $M_{\mu^-j_1j_2} = M_{1}$ and $M_{\mu^-j_1j_2} = M_{2}$
is hardly observable, but the fact that $A_{FB} = 0$
at two different values of $M_{\mu^-j_1j_2}$ would be clear evidence
of the production of two  heavy Majorana neutrinos with
just different masses. We showed that at the same time 
the $\cos\theta$-distribution at $M_{\mu^-j_1j_2} = M_{3}$
differs noticeably from those  at
$M_{\mu^-j_1j_2} = M_{1},M_2$.  

We showed also that in all the cases considered by us
the heavy Majorana neutrinos will decay within the detector with different 
displaced vertices of the decays, which would be one additional clear
signature of their production.

To summarise, the forward-backward symmetry/asymmetry indicator of the
nature - Majorana or Dirac -  of a neutral heavy lepton 
produced together with a flavour neutrino in $e^+ - e^-$ collisions is an
alternative powerful tool to probe, e.g., the Majorana nature of
the heavy neutrinos, which are an integral part of the type I seesaw
mechanism of neutrino mass generation, and thus to test
the possibility of lepton charge nonconservation in Nature.

It follows  from our study, in particular, that the
experiments at FCCee
during its first phase of operation have the
remarkable potential not only to discover the heavy neutral leptons associated with the low-scale seesaw mechanism of neutrino mass generation if they have
masses smaller than the $W^\pm$-boson mass, 
but also to probe and provide evidence for their Majorana nature.

We hope the results of the present work can serve as guidance and
  inspiration
for the future experimental searches for the heavy Majorana neutrinos associated
with the low-scale seesaw scenarios of generation of the light
neutrino masses. 

%
\section*{Acknowledgements}
%

T.O.  would like to thank P.S. Bhupal Dev, Keisuke Harigaya, Gordan Krnjaic, and Stephen J. Parke for useful discussions. The work of T.O. is partly supported
by the INFN initiative Theoretical Astroparticle Physics (TAsP). T.O. thanks the Fermilab Theory Division for its hospitality and support through the Summer Visitors’ Program. T.O. also thanks Washington University in St. Louis, the University of Oklahoma, Pennsylvania State University, and Boston University for their hospitality and support during visits where part of this work was carried out.
The work of S.T.P.~was supported in part by the European Union's Horizon Europe research and innovation programme under the Marie Sk\l{}odowska-Curie grant agreement No. 101086085-ASYMMETRY, by the Italian INFN program on Theoretical Astroparticle Physics and by the World Premier
International Research Center Initiative (WPI Initiative, MEXT), Japan.

\vspace{0.8cm}
\bibliographystyle{utcaps_mod} 
\bibliography{bibliography}

\end{document}